\documentclass[prl,twocolumn,amsmath,amssymb,10pt,aps,longbibliography,superscriptaddress,citeautoscript,bibnotes,footinbib]{revtex4-1}

\usepackage{feynmp}
\usepackage{amsmath}
\usepackage{graphicx}
\usepackage{multirow}
\usepackage{latexsym}
\usepackage{textcomp}
\usepackage{verbatim}
\usepackage{color}
\usepackage{bm}
\usepackage{subfigure}
\usepackage{everysel}
\usepackage{keyval}
\usepackage{ragged2e}
\usepackage{dsfont}
\usepackage{amssymb}
\usepackage{enumitem}
\usepackage{pstricks}
\usepackage{mathtools}
\usepackage{braket}
\usepackage{slashed} 
\usepackage[colorlinks=true]{hyperref}

\usepackage{graphicx}
\usepackage{xcolor}
\usepackage{amsmath}
\usepackage{amsfonts}
\usepackage{bbm}

\newcommand{\osum}{{%
    \setbox0\hbox{\circ}%
    \rlap{\hbox to \wd0{\hss\sum\hss}}\box0
}}

\begin{document}

\title{Stable Flatbands, Topology, and Superconductivity of Magic Honeycomb Networks}

\author{Jongjun M. Lee}
\affiliation{Department of Physics, Pohang University of Science and Technology (POSTECH), Pohang 37673, Republic of Korea}

\author{Chenhua Geng}
\affiliation{Institute for Solid State Physics, The University of Tokyo, Kashiwa, Chiba 277-8581, Japan}

\author{Jae Whan Park}
\affiliation{Center for Artificial Low Dimensional Electronic Systems,  Institute for Basic Science (IBS),  Pohang 37673, Korea}

\author{Masaki Oshikawa}
\affiliation{Institute for Solid State Physics, The University of Tokyo, Kashiwa, Chiba 277-8581, Japan}

\author{Sung-Sik Lee}
\affiliation{Department of Physics \& Astronomy, McMaster University, 1280 Main St. W., Hamilton Ontario L85 4M1, Canada}
\affiliation{Perimeter Institute for Theoretical Physics, 31 Caroline ST. N., Waterloo Ontario N2L 2Y5, Canada}

\author{Han Woong Yeom}
\affiliation{Center for Artificial Low Dimensional Electronic Systems,  Institute for Basic Science (IBS),  Pohang 37673, Korea}
\affiliation{Department of Physics, Pohang University of Science and Technology (POSTECH), Pohang 37673, Republic of Korea}

\author{Gil Young Cho}
\thanks{Electronic Address: gilyoungcho@postech.ac.kr}
\affiliation{Department of Physics, Pohang University of Science and Technology (POSTECH), Pohang 37673, Republic of Korea}

\date{\today}
 
\begin{abstract} 
We propose a new principle to realize flatbands which are robust in
real materials, based on a network superstructure of one-dimensional segments.
This mechanism is naturally realized in the nearly commensurate
charge-density wave of 1T-TaS${}_2$ with the honeycomb network
of conducting domain walls, and the resulting flatband
can naturally explain the enhanced superconductivity.
We also show that corner states, which are a hallmark of the
higher-order topological insulators, appear in the network superstructure.
\end{abstract}

\maketitle
Band theory of electrons has gone through a renaissance in recent
years, especially concerning its topological nature~\cite{Top1, Top2}.
Band structures are also an important starting point to understand
strongly correlated systems.
In particular, when the Fermi level lies in a flatband,
the correlation effects become dominant, leading to many interesting
physics including superconductivity and magnetism.
A few theoretical principles to realize flatbands are known, most notably
the chiral (sublattice) symmetry and an imbalance between the
number of sublattice sites~\cite{ShimaAoki1993, Flat-Chiral}.
However, once further nearest-neighbor hoppings, which generally
exist in real materials, are included, the chiral symmetry is lost
and the flatbands in the simplified model acquire a substantial curvature.
This is a major reason
why it has been difficult to realize flatbands in real
materials and observe resulting strong correlation effects experimentally,
with few exceptions~\cite{cao2018unconventional,cao2018correlated,bistritzer2011moire}.

In this Letter, we propose a novel general principle to realize
flatbands based on a network of one-dimensional segments.
The resulting flatbands are protected by the combination of crystal and time-reversal symmetries,
and are robust even in the presence of the further neighbor hoppings,
as long as the effective hopping range is shorter than the segment length.
We argue that this mechanism is naturally realized in
the nearly commensurate charge-density wave (NC-CDW) phase
of 1T-TaS${}_2$, in which
the domain walls play the role of the one-dimensional metallic segments~\cite{Park-Prep}.
The robust flatbands ensured by the new principle
give a natural explanation of the observed
superconductivity, which is strongly reminiscent of the moir\'e physics~\cite{cao2018unconventional,cao2018correlated,bistritzer2011moire} of twisted bilayer graphene.
Moreover,  the network superstructure can also
lead to corner states, i.e., a higher-order topological insulator. 

Experimentally, 1T-TaS${}_2$ has a rich phase diagram of CDW orders~\cite{wilson1975charge} and superconductivity (SC)~\cite{sipos2008mott, yu2015gate, PhysRevB.94.045131}, which are accessible by tuning temperature, pressure, and doping. There are three distinct regimes: commensurate CDW (C-CDW), NC-CDW, and incommensurate CDW. The C-CDW phase has a long-ranged charge ordering and appears at the lowest temperature with the ambient pressure. This state is the correlation-driven Mott insulator~\cite{fazekas1980charge, PLee-1, PLee-2}. When the C-CDW ordering is slightly suppressed by pressure or doping, then the domain walls appear in between the locally charge-ordered domains, i.e., it enters into the NC-CDW state. If the pressure or doping increase further, the SC emerges from this NC-CDW state. Note that a similar phase diagram is obtained in another CDW material, namely 1T-TiSe${}_2$~\cite{li2016controlling}. This suggests that the NC-CDW~\cite{sipos2008mott, yu2015gate, PhysRevB.94.045131, li2016controlling,chen2019discommensuration} is somehow essential for realizing SC in these CDW materials though how this actually happens has been unclear.  
We will point out that a honeycomb network of metallic domain walls
in the NC-CDW phase~\cite{NC-CDW-Xray, Park-Prep, wu1989hexagonal}
hosts a series of robust flatbands ensured by the new principle, giving
a natural explanation for the observed SC.

\begin{figure}[t] 
\centering{ \includegraphics[width=1.0\columnwidth]{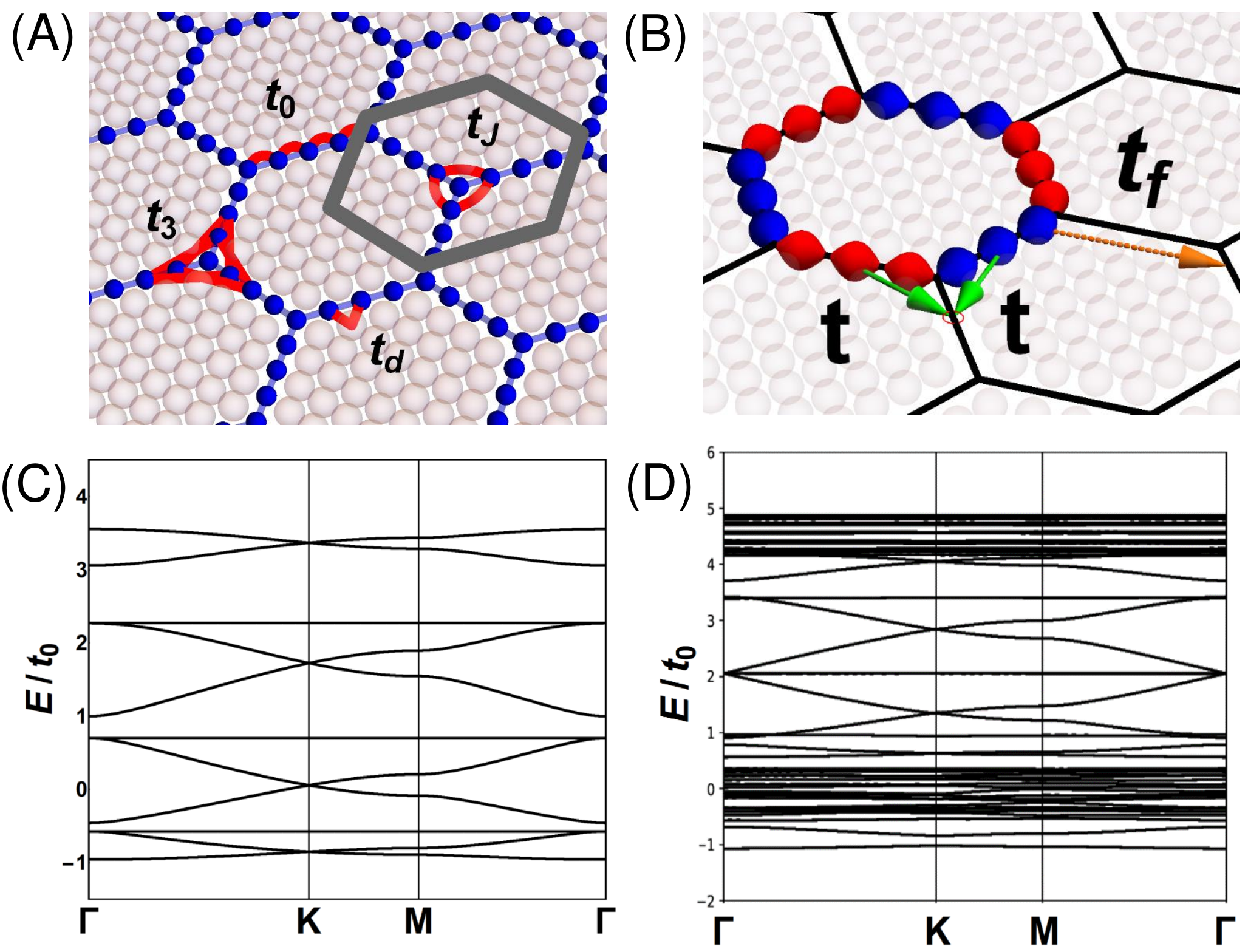} }
\caption{ 
(Color online) (A) Network of 1D metals. In Eq.\eqref{TB}, only the sites along the straight blue lines are included. The other sites represent the insulating domain sites. The grey line represents the unit cell. (B) The states inside the flatbands. Blue and red colored regions have the opposite signs and a destructive interference appears for any hopping shorter than the link length (e.g., $t$). (C) Band structure of Eq.\eqref{TB}. (D) Band structure including the domain sites. 
}\label{Fig1}
\end{figure}

\textbf{1. Model and Flat Bands}: A recent STM study~\cite{Park-Prep} combined with DFT calculation provided an unprecedented detail of the electronic structure of the network, where the metallic nature of the domain walls and trijunctions is clearly exposed. Similar structure emerges in, e.g., helium mixture absorbed to graphite~\cite{Morishita2019}. Here we consider a minimal tight-binding model, which consists of the low-energy modes inside the domain walls; see Fig. \ref{Fig1}(A). 
\begin{align}
H_{0} = -t_{0} \sum_{\langle \bm{r}, \bm{r}' \rangle}  c^{\dagger}_{\bm{r}, \sigma}  c_{\bm{r}', \sigma}  -t_J \sum_{\{\bm{r}, \bm{r}' \} \in J}  c^{\dagger}_{\bm{r}, \sigma}  c_{\bm{r}', \sigma} + H.c., 
\label{TB}
\end{align}
where the second sum over $\{\bm{r}, \bm{r}'\} \in J$ runs over the sites around the junctions $J$ and $t_J \leq t_0$. Here $\sigma$ represents the spin and the sum over it is assumed. Here we assumed the spin rotational symmetry for simplicity. The model also has time-reversal symmetry $\mathcal{T}$, and crystalline $D_6$ symmetry, which are the symmetries observed in the STM experiment of 1T-TaS${}_2$~\cite{Park-Prep}. Diagonalizing Eq.\eqref{TB}, we find a cascade of flatbands, Dirac and quadratic band crossings; see Fig. \ref{Fig1}(C). On top of this, a three-component spin-$1$ Dirac fermion~\cite{dora2011lattice} can appear when $t_J/t_0 \to 0$. The number of the flatbands is proportional to the number of the sites between the junctions. The topological band crossings are protected by symmetries. On the other hand, the flat dispersion cannot be generally protected because it requires infinitely many parameters to be tuned. Hence, they are generically fragile, e.g., against the second neighbor hoppings~\cite{SI}. Despite of the fragile nature, flatbands are an ideal stage to realize correlation-dominated physics,
such as ferromagnetism, and thus have been studied
vigorously~\cite{Leykam-flatbands2018}.
A well-known mechanism which gives rise to flatbands
is an imbalance between sublattice sites in bipartite
lattices. This was applied~~\cite{ShimaAoki1993}
to hyperhoneycomb systems which have
some similarities with the systems we study in this paper.
However, such flatbands can be generically removed by
inclusion of short-ranged further neighbor hoppings, which exist in real
materials.



Remarkably, our flatbands from the network defy this standard phenomenology and are stable against $D_6$-symmetric local perturbations. For example, addition of the third-neighbor hoppings $t_3$ near the nodes do not disperse the flatbands. We can even include the insulating electronic states from the domain area (described by $H_{\text{dom}}$); see Fig. \ref{Fig1}(A). The full Hamiltonian is now $H = H_{0} + H_{\text{dom}} + H_{\text{coup}}$ with  
\begin{align}
H_{coup} = -t_{d} \sum_{\langle \bm{r}, \bm{r}' \rangle}c_{\bm{r}, \sigma}^{\dagger} d_{\bm{r}', \sigma} + H.c., \nonumber
\end{align}
where $c_{\bm{r}, \sigma}$ ($d_{\bm{r}, \sigma}$) electrons belong to the network (to the domains). Here $H_{\text{dom}}$ is described by the band insulator which has the two energy-split states per site and the different sites are connected by the small hopping $t_d$ [Fig. \ref{Fig1}(A)]. From the perturbative reasoning, we expect that this model includes all the possible symmetric local perturbations. Notably, the flatbands and overall band shapes remain almost intact inside the gap [Fig.\ref{Fig1}(D)]. This result implies that in NC-CDW 1T-TaS${}_2$, even if the bulk bands from the domain region are included, flatbands inside the gap are almost intact. Finally, the flatbands survive~\cite{SI} against the Rashba spin-orbit coupling. Such stability is absent for other networks~\cite{SI}. In passing, we note that this is consistent with our previous phenomenological approach; see~\cite{SI}. 


To explain the unusual stability, we look into the structure of the wave functions inside the flatbands. We find that those wave functions vanish at the junctions~\cite{SI}. Hence, when the low-energy modes are entirely from the network and only the nearest neighbor hoppings are included, i.e., $t_J =0$ in Eq.\eqref{TB}, the wave function is a standing wave $\psi(l)$ [Fig. \ref{Fig1}(B)]. From such standing waves, we can construct a set of localized states~\cite{PhysRevB.78.125104} which consist of the flatbands: we consider a linear combination around the honeycomb plaquette with a sign oscillation, i.e., $\Psi \sim \sum_a (-1)^a \psi(l)$ for $a \in \{1,2, \cdots 6\}$ labeling the six links around the plaquette; see Fig. \ref{Fig1}(B). Then we see that this state cannot disperse into the neighboring plaquettes because of the destructive interference. Such destructive interference persists as far as the hopping distance is shorter than the length of the wire. Finally, the intrawire further neighbor terms do not alter this conclusion because they affect only the energy and intrawire structure of the standing waves~\cite{SI}. This illustrates the importance of the symmetry and the locality on the stability of the flatbands. We can also understand that the cascade of the flatbands \textit{must} appear because there are many standing wave solutions per wire~\cite{Hosho}. 


We expect the flatbands to be removed either by long-ranged direct hopping across the domains or by breaking symmetries. Indeed, we can show~\cite{SI} that the flatbands are lifted when the symmetries are broken, or when such long-ranged hoppings [e.g., $t_f$ in Fig. \ref{Fig1}(B)] are included. This means that, for the NC-CDW state of 1T-TaS${}_2$, we need a sizable hopping between the two sites apart by $\sim O(80) \r{A}$~\cite{Park-Prep} to remove the flatbands, which is not realistic. This explains why the dispersion of the ``flatbands" are very small but still finite when the domain sites are included. When the time-reversal symmetry is broken, the band touchings are gapped out and it results in dispersive Chern bands~\cite{SI}, which is a natural platform for fractionalization~\cite{PhysRevLett.106.236802}. 



For 1T-TaS${}_2$, we constructed a tight-binding model in the Supplemental Material~\cite{SI}, which fits reasonably well with DFT+U calculation on the domain wall. The tight-binding parameters scale as $1/d^5$~\cite{PhysRevLett.122.036802} where $d$ is the distance between the atomic sites. From this, we find that the cascade of the flatbands emerges~\cite{SI}. Next, we comment on the effect~\cite{Inter1, Inter2, Inter3, Inter4} of the interlayer coupling. The interlayer interaction has been suggested to be important in 1T-TaS${}_2$. However, we remark that there are plenty of experimental data and theory~\cite{Park-Prep, PLee-2} suggesting that the main physics is essentially 2D. For example, the resistivity along the $c$ axis is much larger, e.g., by 500 times~\cite{hambourger1980electronic}, than the intralayer resistivity~\cite{hambourger1980electronic, svetin2017three} and anisotropic 2D charge transfer is observed for the NC-CDW state~\cite{kuhn2019directional}. Further, the SC $T_c$ is almost insensitive to the pressure~\cite{sipos2008mott} and not much affected under the dimensional reduction~\cite{yu2015gate}. Based on these, we focus on the 2D physics here.

\textbf{2. Superconducting states}: Having established the stability of the flatbands, we now discuss the many-body physics when the Fermi level is near one of the flatbands. Such a system is unstable toward various particle-hole and particle-particle channels. However, guided by experiments, we mainly focus on superconductivity in this paper. 

\begin{figure}[t] 
\centering{ \includegraphics[width=1.0\columnwidth]{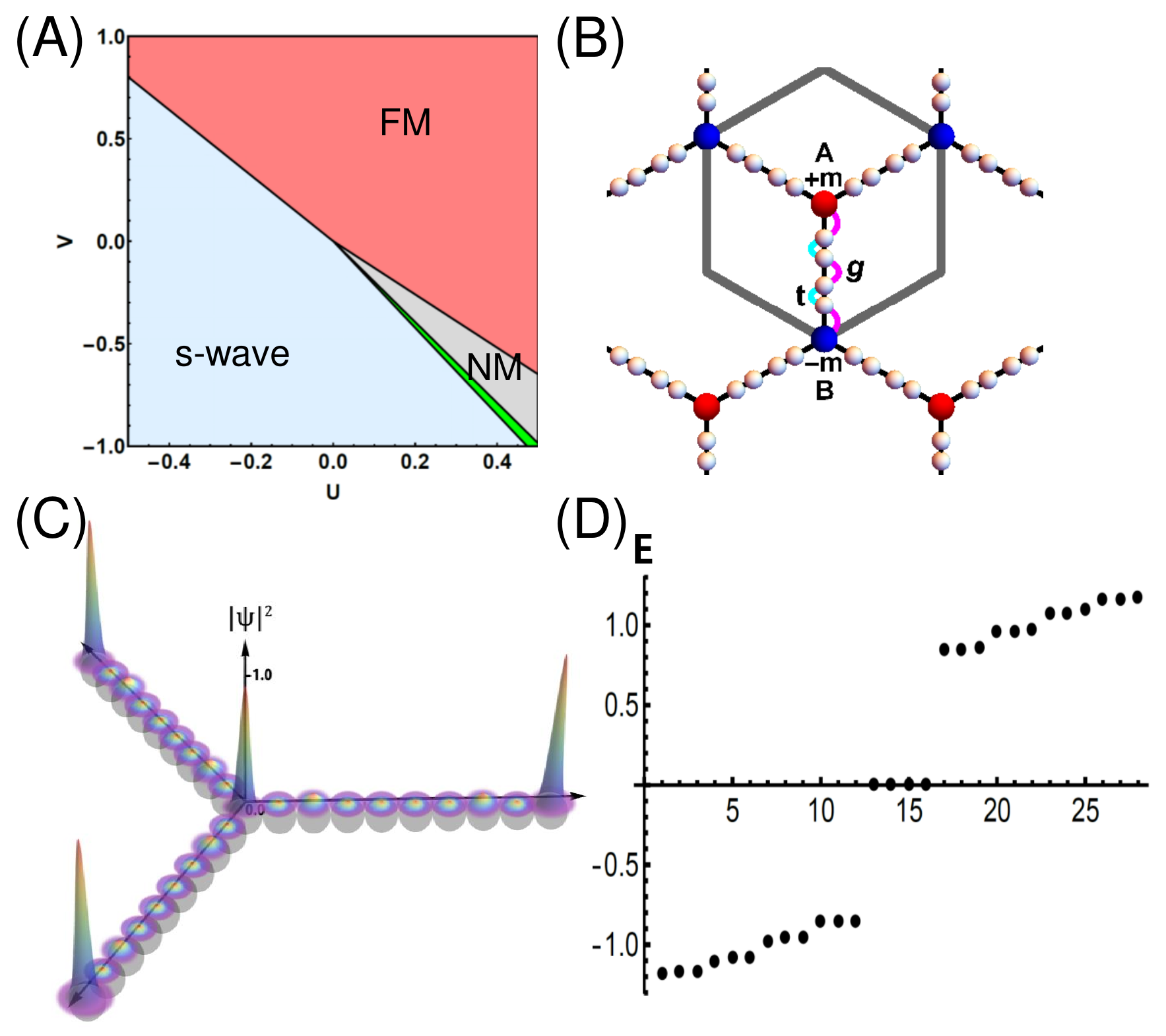} }
\caption{ 
(A) Phase diagram. ``$s$-wave" represents an $s$-wave SC, and ``FM" (``NM") represent ferromagnetic (normal metal) state. The green region is a ($d+id$)-wave SC. (B) Pictorial representation of the model for the higher-order topology. Here $|t|<|g|$ supports the 0D state at the junctions and the corner at the boundary. The grey line represents the unit cell. (C) Density distribution of four in-gap modes of (D). (D) Energy eigenvalues for a single unit cell with open boundary condition for $t/g=0.1$. 
}\label{Fig2}
\end{figure}

First, we perform the simplest BCS mean-field theory with the phenomenological interaction  
\begin{align}
H_{int} = U \sum_{\bm{r}} n_{\bm{r}}^2 + V \sum_{\langle \bm{r}, \bm{r}' \rangle} n_{\bm{r}} n_{\bm{r}'}. \label{M-Int}
\end{align}
Projecting to the BCS channel, we obtain~\cite{SI} 
\begin{align}
H_{int}  \to H_{\text{BCS}} = \sum_{l \in 2\mathbb{Z}} g_l  \sum_{\bm{p}, \bm{k}} \hat{\triangle}_{l; \bm{p}}^{\dagger} \cdot \hat{\triangle}_{l; \bm{k}}, \label{BCS-Int}
\end{align}
where $g_l$ is the interaction strength along the pairing channel of the angular momentum $l$, which we compute numerically. $\hat{\triangle}_{l;\bm{p}}$ is the corresponding pairing order parameter. Note that, within the mean-field decomposition of Eq.\eqref{M-Int}, only the spin-singlet sector appears. Below we consider only the $s$-wave and $(d+id)$-wave pairing channels, i.e., $g_0$ and $g_2=g_{-2}$ of Eq.\eqref{BCS-Int}.  Higher angular momentum pairing channels ($|l|>2$) will belong to the same representations of $s$- or $(d\pm id)$-wave pairing channels in the honeycomb symmetry. The magic of the flatbands appears when the gap equation is solved. 
\begin{align}
\frac{1}{g_l} \sim \int_{\text{BZ}} \frac{d^2 k ~ |F_l (\bm{k})|^2}{\sqrt{(\epsilon_k - \mu)^2 + |F_l (\bm{k})\triangle_l|^2}} \sim \frac{1}{|\triangle_l|}, \nonumber   
\end{align}
where $F_l (\bm{k})$ is the form factor~\cite{SI}, e.g., $F_2 (\bm{k}) \to (k_x + ik_y)^2$ for $|\bm{k}| \ll 1$ and $\epsilon_{\bm{k}}=\mu$. Hence, the gap is linearly enhanced, i.e., $|\triangle_l | \sim |g_l|$ (if $g_l <0$), instead of the standard exponential suppression $\sim \exp(-1/|g_l |\nu_l)$. Because there is no other scale, the mean-field energy of the SC is linear in $g_l$ and so is the $T_c$~\cite{SI}. Hence, the honeycomb network strongly enhances the SC $T_c$. The phase diagram for one flatband is in Fig.\ref{Fig2}(A).
Thus we conclude that, within the mean-field theory,
the $s$-wave SC is strongly enhanced when $U$ is attractive~\cite{Flat-SC}.
On the repulsive interaction side, the system exhibits ferromagnetism.  
While the repulsive $U$ opens up a small window for the $(d\pm id)$-channel
pairing, the dominance of the $s$-wave SC is a characteristic feature
of the present flat-band system.~\cite{SI}. In passing, the BdG fermion spectrum is also computed~\cite{SI}.

Although the mean-field theory ignores the fluctuation, it
is a good starting point for revealing possible phases and phase diagram.
In fact, the mean-field theory qualitatively agrees with
a rigorous result on flat-band ferromagnetism~\cite{PhysRevLett.69.1608}
with a repulsive $U$, and 
our findings in this Letter are consistent with
recent numerical studies in other
flat-band systems~\cite{Flat-SC, n2, n3, n4}.

Let us comment on the interaction Eq.\eqref{M-Int}. The strong electron-phonon coupling in 1T-TaS${}_{2}$ is known as the driving force for the formation of the CDW states~\cite{Liu2009}. However, after the formation of the CDW clusters through the electron-phonon coupling, the soft phonons are naturally hardened~\cite{Lazar2015}. In particular, the domain wall in the NC-CDW state is also structurally reconstructed by forming its own CDW clusters from electron-phonon coupling,~\cite{Park-Prep} which will reduce the coupling more. Hence, we expect that the electron-phonon coupling becomes inactive for the physics within the NC-CDW phase. Nevertheless, the phonon-electron interaction was essential in forming the parent CDW domains and the domain wall network, and hence its effect is already encoded the model Eq.\eqref{TB}. Next, it is well known that the coupling of electrons and (optical or gapped) phonons effectively plays the same role as the attractive $U$, which favors the $s$-wave SC.\cite{SI} This will effectively renormalize $U$ in Eq.\eqref{M-Int} toward negative. Hence, explicitly including the effects of these phonons does not alter the dominance of the $s$-wave SC.\cite{SI} On the other hand, the long-range Coulomb interaction will be efficiently screened due to the large density of states of flatbands and it will be rendered into a short-ranged interaction, which can be effectively encoded by Eq.\eqref{M-Int}. 


It is instructive to consider the strong-coupling limit,~\cite{wu2019coupled} where the interaction is bigger than the hopping integrals. For this, we start from the decoupled \textit{strongly correlated} wires, each of which is described by a Tomonaga-Luttinger liquid (TLL)
\begin{align}
H =  \sum_{a, \sigma = c/s} \int dl ~  \frac{v_\sigma}{2} \Big[\frac{1}{K_\sigma} (\partial_l \theta_{a,\sigma} )^2 + K_\sigma (\partial_l \phi_{a,\sigma})^2\Big], \nonumber 
\end{align}
where the Luttinger parameters $\{K_c, K_s \}$ capture the correlation nonperturbatively~\cite{fradkin2013field}. This is the correlated version of the scattering problem~\cite{Park-Prep}, which faithfully reproduced the band structures. 

The two-dimensional SC can be preempted by the spin gap~\cite{fradkin2013field, kivelson1998electronic} in each wire. Once the spin gap forms, the low-energy physics of each wire is described by a single-component TLL
of $\theta_{a,c}$ describing the fluctuating SC pairs~\cite{fradkin2013field}.
Since the SC pair is bosonic, the junction of three TLLs at each vertex
of the honeycomb network corresponds to the bosonic $Y$ junction~~\cite{PhysRevLett.100.140402}, rather than the fermionic one~~\cite{PhysRevLett.91.206403, oshikawa2006junctions} where the fermion statistics plays an important role.
When each wire has sufficient attraction, i.e., $K_c \leq 1$~\cite{fradkin2013field}, then the interwire coupling between the SC fluctuations, namely, the Josephson coupling $J$, becomes relevant~\cite{fradkin2013field}. Only interested in the pattern of the phases, we note that the problem is symmetrically equivalent to the $XY$ model on the kagome lattice, while leaving the full quasi-1D treatment~\cite{kivelson1998electronic} to the future study
\begin{align}
H_{eff} = J \sum_{\langle i, j \rangle} \Big[e^{ \sqrt{2\pi} i(\theta_{i,c}-\theta_{j,c})} + h.c.\Big] \label{SC}
\end{align}
Depending on the sign of $J$, either the 2D $s$-wave or $(d \pm id)$-wave SCs can emerge. When $J>0$, then the so-called $\sqrt{3}\times \sqrt{3}$ order appears~\cite{PhysRevB.48.9539}, which translates as the $(d\pm id)$-SC. If $J<0$, the conventional $s$-wave SC emerges. When the repulsive-$U$ dominates the junction region and the region becomes Mott insulating~\cite{berg2009striped}, $J>0$ can appear. From this strong-coupling limit, we can learn how the $2k_F$-density wave of 1d wires competing with SC is suppressed. For the generic filling of each wire, the momentum $k_F$ will not be commensurate with the wire length $L$, i.e., $\phi_L = k_F L$ is not a rational number. This frustrates the phases of the density waves and thus their true two-dimensional order is strongly suppressed to develop. Because the density waves are the main competitors of the SC in one dimension, this gives a natural favor on the SC. 

The domain wall states of 1T-TaS${}_2$ presumably experience small correlation effect and the junction regions are quite metallic from the STM data~\cite{Park-Prep}. Furthermore, no magnetism is observed in experiments, i.e., interaction is dominantly attractive.  Furthermore, the SC in 1T-TaS${}_2$ is experimentally observed for a broad range of the parameters~\cite{sipos2008mott, yu2015gate}. When compared with Fig.\ref{Fig2}(B), the SC is consistent with the $s$ wave. These suggest that the SC state of 1T-TaS${}_2$ is likely an $s$-wave SC. 

\textbf{3. Higher-order topology}: When the filling per wire is commensurate, then the network can develop its own CDW order and become an insulator. We highlight here that the insulating network hosts an interesting possibility, namely emergent corner states which are akin to those of the 2d higher-order topological insulators~\cite{Benalcazar61, HOTI_3, HOTI_4}. We illustrate this on the half-filled spinless fermion model, whose generalization to the spinful systems is straightforward. In such wires, the dimerization is expected, which manifests as the staggered nearest-neighbor hopping parameters; see Fig.~\ref{Fig2}(B). To clearly expose the corner states, we perform a finite-size calculation on a single unit cell of the network. In the spectrum Fig.~\ref{Fig2}(D), we see four in-gap modes, which are localized at the trijunction and the boundary Fig.~\ref{Fig2}(C). Those states are protected by crystal symmetries. We can also show that the above system similarly supports an in-gap mode at the corner of the edge. See Supplemental Material~\cite{SI} where the junction of two gapped domain walls supports a corner mode protected by a reflection symmetry. Such 0D states protected by crystalline symmetries are the hallmarks of higher-order topology.

The above higher-order topology of gapped domain wall networks may be relevant to various materials with similar structures such as C-CDW 1T-TaS${}_2$~\cite{cho2017correlated} and twin boundary networks of MoSe${}_2$ grown on MoS${}_2$~\cite{ACS}. Spectacularly, it has been noted~\cite{cho2017correlated, PhysRevLett.122.036802} that the insulating domain walls of C-CDW 1T-TaS${}_2$ have precisely the same structure as the dimerization, and they form junctions and networks. We believe that the confirmation of the higher-order topology in these systems will be an extremely interesting future problem, given that there is no concrete experimental demonstration of the 2D higher-order topology so far.

\textbf{4. Conclusions}: We considered the electronic structure of a conducting honeycomb network, where we uncovered the emergence of the cascade of the flatbands that are stable against various local perturbations. Compared to the previous studies~\cite{Leykam-flatbands2018,barreteau2017bird, sun2016half, zheng2014exotic, maruyama2016coexistence, ShimaAoki1993, PhysRevB.78.125104}, our work reveals that the flatbands can emerge in much broader sets of the models beyond the models with chiral symmetry or only nearest-neighbor hopping terms. We also demonstrated that the domain wall network is an ideal place to find diverse topological band structures: topological band crossings, and higher-order corner states. We find that the robust flatbands and network geometry can explain the coexistence of SC and CDW states in the network materials, which goes beyond the previous Landau-Ginzburg theory~\cite{chen2019discommensuration}, which is blind to the emergent electronics from the network.

The signature of the network and its role in SC can be detected in various experiments in 1T-TaS${}_2$. First, the combination of our STM data, DFT calculation,~\cite{Park-Prep} and the result of our current paper already points strongly toward the existence of the flatbands. In particular, the scanning tunneling spectroscopy~\cite{Park-Prep} showed that the band gap is filled, which again supports the emergence of cascade of band structures. In particular, one nearly flatband is observed right below the chemical potential in the currently available photoemission data~\cite{ARPES1, ARPES2, ARPES3} though further investigation will be desirable. Second, magnetotransport and oscillations can provide the information about the primary conducting and SC channels. They have been applied in the small twist-angle bilayer graphene~\cite{rickhaus2018transport, efimkin2018helical} and textured superconducting states of 1T-TiSe${}_2$~\cite{li2016controlling, chen2019discommensuration}. Also the flatbands show a few characteristic behaviors in thermodynamic quantities, which we summarize in the Supplemental Material~\cite{SI}.
It would be also interesting to perform numerical calculation on
our network model with interactions, to confirm our predictions.


\acknowledgements
We thank Peter Abbamonte, Ehud Altman, Liang Fu, Hiroshi Fukuyama, Jung-Hoon Han, Tim Hsieh, Eun-Ah Kim, Yong Baek Kim, Jong Hwan Kim, Hae Young Kee, Tae Hwan Kim, Sungbin Lee, Ivar Martin, Masashi Morishita, Adrian Po, Youngwoo Son, Dam T. Son, Ashvin Vishwanath, and Mike Zaletel for helpful discussion. The research of S.~L. was supported by NSERC. Research at the Perimeter Institute is supported in part by the Government of Canada through Industry Canada, and by the Province of Ontario through the Ministry of Research and Information.
C.~G. and M.~O. were supported in part by MEXT/JSPS KAKENHI Grant No. JP18H03686 and No. JP17H06462. J.W. P. and H.W.Y are supported by the Insitute for Basic Science (IBS-R014-D1). G.Y.C thanks for the support of the Visiting Fellowship at the Perimeter Institute. Part of this work is done during the winter program ``New Approaches to Strongly Correlated Quantum Systems'' at Aspen Center for Physics, supported by the US National Science Foundation Grant No. NSF PHY-1607611,
and at the Kavli Institute for Theoretical Physics, UC Santa Barbara, supported by NSF PHY-1748958. This work was supported by the National Research Foundation of Korea(NRF) grant funded by the Korea government(MSIT) (No. 2020R1C1C1006048).
After the completion of this work, we became aware of independent work on decorated star lattices~\cite{mizoguchi2019flat} where flatbands and higher-order topology are also discussed.  We thank Hosho Katsura for several useful comments and bringing our attention to this recent independent paper~\cite{mizoguchi2019flat} and inform us about Ref.~\cite{Hosho}, where the energy of the flatbands is shown to be related to the energy of a single wire subject under the open boundary condition. 

\bibliography{Network_0310}

\end{document}


\title{Supplemental Information for ``Stable Flat Bands, Topology, and Superconductivity of Magic Honeycomb Network"}

\author{Jongjun M. Lee}
\affiliation{Department of Physics, Pohang University of Science and Technology (POSTECH), Pohang 37673, Republic of Korea}

\author{Chenhua Geng}
\affiliation{Institute for Solid State Physics, The University of Tokyo, Kashiwa, Chiba 277-8581, Japan}

\author{Jae Whan Park}
\affiliation{Center for Artificial Low Dimensional Electronic Systems,  Institute for Basic Science (IBS),  Pohang 37673, Korea}

\author{Masaki Oshikawa}
\affiliation{Institute for Solid State Physics, The University of Tokyo, Kashiwa, Chiba 277-8581, Japan}

\author{Sung-Sik Lee}
\affiliation{Department of Physics \& Astronomy, McMaster University, 1280 Main St. W., Hamilton ON L85 4M1, Canada}
\affiliation{Perimeter Institute for Theoretical Physics, 31 Caroline ST. N., Waterloo ON N2L 2Y5, Canada}

\author{Han Woong Yeom}
\affiliation{Center for Artificial Low Dimensional Electronic Systems,  Institute for Basic Science (IBS),  Pohang 37673, Korea}
\affiliation{Department of Physics, Pohang University of Science and Technology (POSTECH), Pohang 37673, Republic of Korea}

\author{Gil Young Cho}
\thanks{Electronic Address: gilyoungcho@postech.ac.kr}
\affiliation{Department of Physics, Pohang University of Science and Technology (POSTECH), Pohang 37673, Republic of Korea}

\date{\today}
\maketitle
\tableofcontents
\appendix
\section{Details of Scattering Description of Honeycomb Network}
Here we summarize the theory part of our previous work\cite{Park-Prep}. Specifically, we will introduce the scattering description of the honeycomb network, which reproduces the key structure of the tight-binding model in the main text. 

The STM experiment\cite{Park-Prep} essentially found that the domains of the nearly-commensurate charge-density wave form a regular honeycomb lattice, and thus the domain walls are the links of this regular honeycomb lattice. Furthermore, the domain walls trap finite local density of states near the Fermi level. We note that the domain walls are generically expected to trap some in-gap modes due to the topological solitonic modes (though these modes may appear away from the Fermi level.). Motivated from these findings, we consider a regular array of one-dimensional metals living on the links of a honeycomb lattice. Similar network models of one-dimensional metals have been studied in the context of quantum Hall plateau transition, known as ``Chalker-Coddington model"\cite{chalker1988percolation}, and also in the twisted bilayer graphene at a small twisting angle\cite{efimkin2018helical}. 

\subsection{Model}
\begin{figure}[h]
\centering
\includegraphics[scale=0.4]{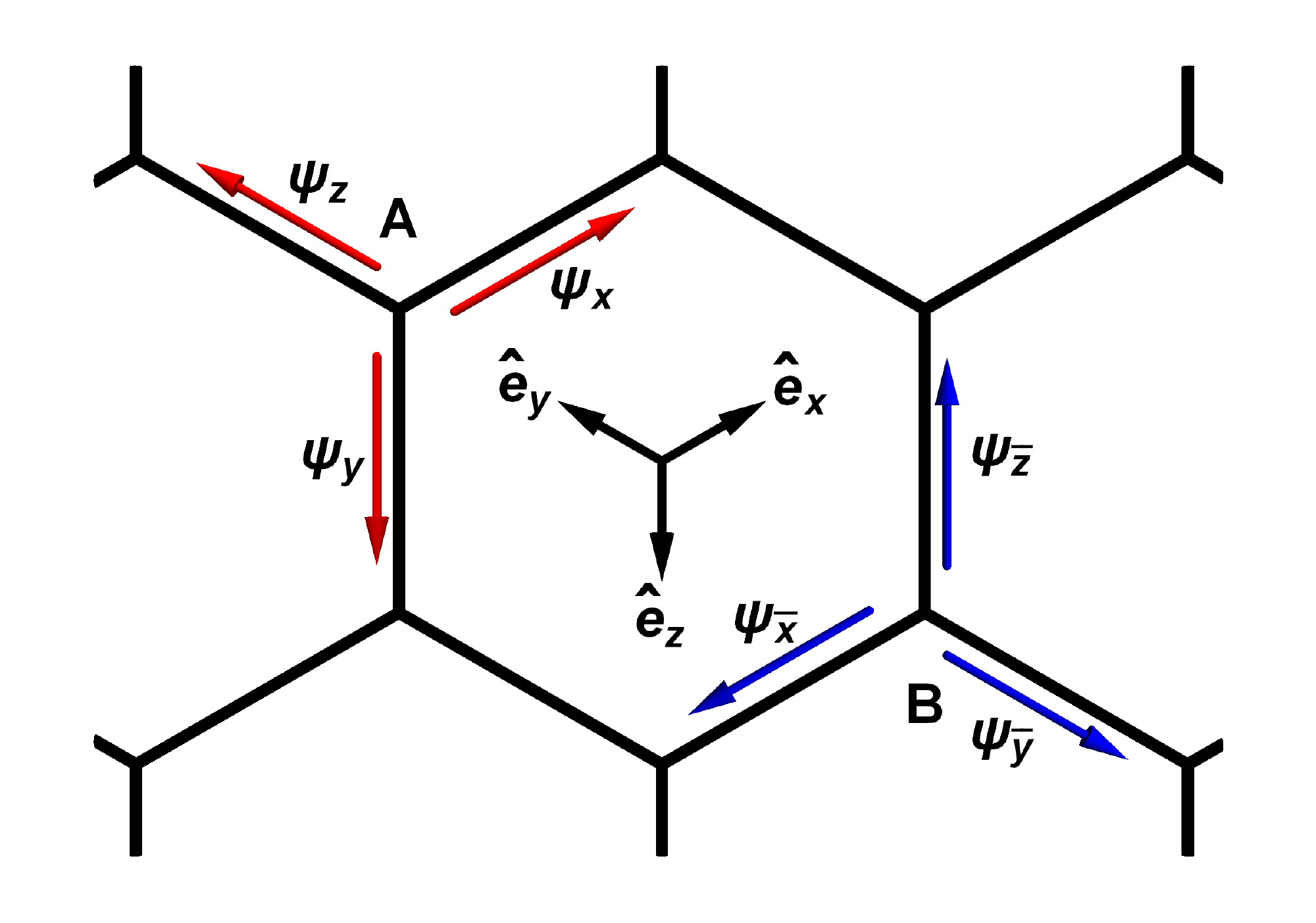}
\caption{Pictorial Representation of Network Model.}
\label{sfig1}
\end{figure}

To capture the physics of the network, we introduce the two wavefunctions on the links of the honeycomb network: $\psi_a$ and $\psi_{\bar{a}}$. Here $\psi_a$ represent the chiral mode propagating from an A-sublattice (of the network) to a B-sublattice (of the network) and $\psi_{\bar{a}}$ for the mode propagating from a B-sublattice to an A-sublattice. See Fig. \ref{sfig1}. 

Hence, we can associate $\psi_a$ to the node at an A-sublattice and $\psi_{\bar{a}}$ to the node at a B-sublattice, i.e., $\psi_{a}$ is an out-going mode along the link $a= x,y,z$ from an A-sublattice, and $\psi_{\bar{a}}$ is an out-going mode along the link $a= x,y,z$ from a B-sublattice (See Fig. \ref{sfig1}). 

We further assume that these modes propagate ballistically within each link and scatter only at the nodes of the honeycomb lattice. We further assume that there are six-fold rotation $C_6$, mirror $R_x$, and $\mathcal{T}$ symmetries, and the scattering between the modes respects the symmetries. (As in the main text, we assume SU(2) spin rotational symmetry and suppress the spin indices here.)

With all of these in hand, we can write down the scattering problem at an A-sublattice.   
\begin{align}
\left[
\begin{array}{c}
\psi_x (\bm{R}) \\
\psi_y (\bm{R}) \\ 
\psi_z (\bm{R}) 
\end{array} \right] = e^{-i\frac{E}{v_F \hbar} L} \cdot \hat{T}_A \cdot \left[
\begin{array}{c}
\psi_{\bar{x}} (\bm{R} +\hat{e}_x) \\
\psi_{\bar{y}}  (\bm{R} + \hat{e}_y) \\ 
\psi_{\bar{z}}  (\bm{R} +\hat{e}_z) 
\end{array} \right]
\end{align}
Here, the left-hand side $\psi_a (\bm{R}), a= x,y,z$ represents the out-going modes from the A-sublattice, which is related by a scattering matrix $\hat{T}_A$ to the in-coming modes $\psi_{\bar{a}} (\bm{R}), a = x,y,z$ appearing on the righ-hand side (See Fig. \ref{sfig1}). The additional phase factor $\sim \exp (-i\frac{E}{v_F \hbar} L)$ is the phase accumulated by the incoming modes while it propagates coherently from the neighboring B-sublattices to the A-sublattice at $\bm{R}$. Here $v_F$ is the Fermi velocity within the one-dimensional metal, which is expected to be similar to that of the bulk electron, and $L$ is the length of the link. The scattering matrix $\hat{T}_A$ is fixed by the combination of the crystal symmetries and $\mathcal{T}$-symmetry. With the unitarity of the scattering matrix, we find 
\begin{align}
\hat{T}_A = e^{i\chi_A} \left[
\begin{array}{ccc}
T_A & t_A & t_A \\
t_A & T_A & t_A \\ 
t_A & t_A & T_A  
\end{array} \right], ~~ |T_A| \in \Big[\frac{1}{3}, ~1\Big], ~ t_A   = e^{i\phi_A} \sqrt{\frac{1-|T_A|^2}{2}},  
\end{align}
with $\phi_A = \cos^{-1}(\frac{|t_A|}{2|T_A|})$. Similary we have the following scattering problem at the B-sublattice.
\begin{align}
\left[
\begin{array}{c}
\psi_{\bar{x}} (\bm{R}) \\
\psi_{\bar{y}} (\bm{R}) \\ 
\psi_{\bar{z}} (\bm{R}) 
\end{array} \right] = e^{-i\frac{E}{v_F \hbar} L} \cdot \hat{T}_B \cdot \left[
\begin{array}{c}
\psi_{x} (\bm{R} -\hat{e}_x) \\
\psi_{y}  (\bm{R} - \hat{e}_y) \\ 
\psi_{z}  (\bm{R} -\hat{e}_z) 
\end{array} \right],
\end{align}
where $\hat{T}_B=\hat{T}_A$ by the crystal symmetries. Now we can perform the Fourier transformation and solve these scattering problems. On performing the Fourier transformation, we find 
\begin{align}
\bm{\Psi}_{\bm{q}} = e^{-i\frac{E_{\bm{q}}}{v_F \hbar} L} \hat{T}_{\bm{q}} \cdot \bm{\Psi}_{\bm{q}}, ~~ \bm{\Psi}_{\bm{q}} = \left[
\begin{array}{c}
\psi_{x} (\bm{q}) \\
\psi_{y} (\bm{q}) \\ 
\psi_{z} (\bm{q}) \\
\psi_{\bar{x}} (\bm{q}) \\
\psi_{\bar{y}} (\bm{q}) \\ 
\psi_{\bar{z}} (\bm{q})  
\end{array} \right], ~~ 
\hat{T}_{\bm{q}} = \left[
\begin{array}{cc}
0 & \hat{T}_A \cdot \hat{V}_{\bm{q}} \\ 
\hat{T}_B \cdot \hat{V}_{\bm{q}}^* & 0  
\end{array} \right],
\label{Energy}
\end{align}
where $\hat{V}_{\bm{q}} = $ diag $[\exp (i\bm{q}\cdot \hat{e}_x), ~\exp (i\bm{q}\cdot \hat{e}_y), ~\exp (i\bm{q}\cdot \hat{e}_z)]$. Hence, the energy spectrum can be obtained by diagonalizing $\hat{T}_{\bm{q}}$, which is unitary. In terms of the eigenvalues $e^{i \epsilon_{j} (\bm{q})}, j= 1, 2, \cdots 6$ of $\hat{T}_{\bm{q}}$, we have 
\begin{align}
E_{j, \bm{q}}^n = 2\pi \frac{v_F \hbar}{L} n + \frac{v_F \hbar}{L} \epsilon_{j} (\bm{q}), ~~j = 1,2, \cdots 6
\end{align}
Here $n \in \mathbb{Z}$ and thus the minibands are repeating in the energy in period of $2\pi \frac{v_F \hbar}{L}$. Mathematically this ambiguity in $n$ originates from the ambiguity of $\epsilon_j (\bm{q})$ by $2\pi$ appearing in the eigenvalues $e^{i \epsilon_{j} (\bm{q})}, j= 1, 2, \cdots 6$. Physically this repetition can be traced back to the excitation energy of the microscopic one-dimensional modes with the same momentum $\bm{q}$, i.e., for a given $\bm{q}$, there are different one-dimensional modes with energy $2\pi \frac{v_F \hbar}{L} n, ~ n\in \mathbb{Z}$. Thus we expect that this repetition will fill up within the band width of the original parent 1d band. Indeed, this is the band structure that we find from the tight-binding problem, where the certain unit structure repeats in energy.

\subsection{Spectrum of Network}
\begin{figure}[h]
\centering
\includegraphics[scale=0.5]{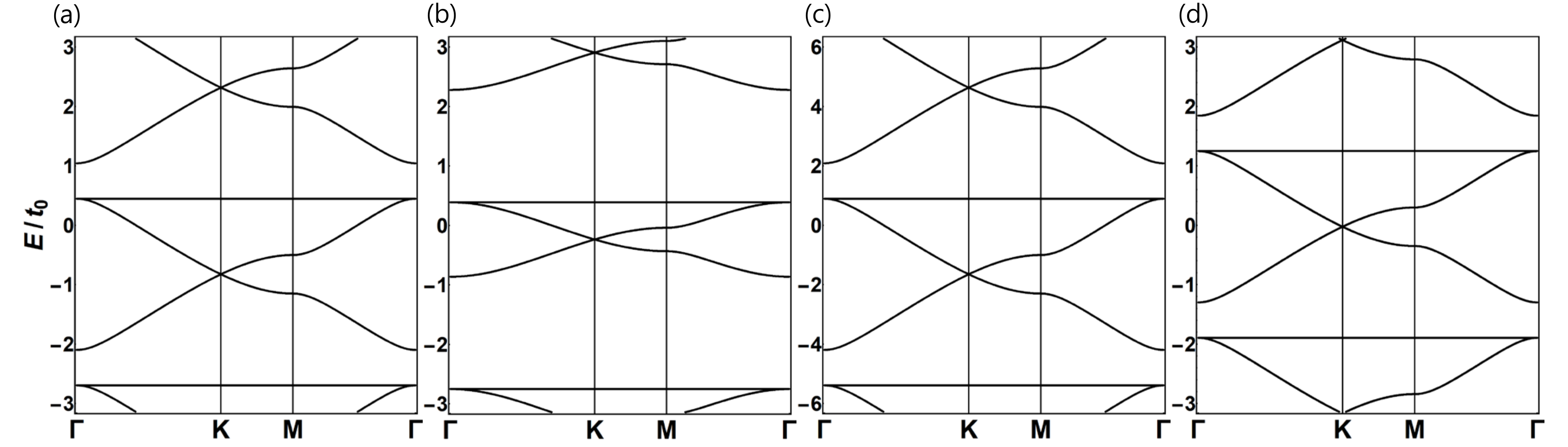}
\caption{Band structure of the network model for several different parameters. (a) $t=0.43$, $v_F \hbar /L =1$, $\chi=0.0$ , (b) $t=0.83$, $v_F \hbar /L =1$, $\chi=0.0$ , (c) $t=0.43$, $v_F \hbar /L =2$, $\chi=0.0$ , (d) $t=0.43$, $v_F \hbar /L =1$, $\chi=0.8$ }
\label{sfig2}
\end{figure}

Next we analyze the band structure out of this scattering description. As apparent from the Fig \ref{sfig2}, the spectrum features (i) Dirac cones at $K$ and $K'$, (ii) flat bands, and (iii) quadratic band touchings at $\Gamma$-point, which are the features of the tight-binding band structure. 

Now the crucial question is if these features are stable against the symmetric deformation of the scattering matrices. We see that the parameters that we can tune are $\{t_A = t_B =t, v_F, \chi \}$, which determine the scattering amplitudes at the nodes and the phase accumulated by the modes while they travel along the links. We find that the overall band structures remain the same. In particular, the flat bands always survive. See Fig. \ref{sfig2}. 

\section{Flat Band Wavefunctions}
Here we reproduce a few standard phenomenology of the flat bands in our model. These include the zeros of the wavefunction in space and the ``frustration" in the hopping Hamiltonians, which is what actually happens in a Lieb lattice. In particular, the former guarantees the existence of a single flat band. In the main text, we go beyond the standard analysis and show the appearance of many stable flat bands. 

\subsection{Presence of Zeros at Tri-junctions}
We first demonstrate that the wavefunction of the flat bands has zeros at the tri-junctions, i.e., nodes of the honeycomb network. For this, we plot out $|\psi_{\bm{q}}(\bm{R})|^2$ at the tri-junction site $\bm{R}$ along a high-symmetry cut in the momentum space, where $\psi_{\bm{q}}(\bm{r})$ is the Bloch function of the flat bands at the momentum $\bm{q}$ and the site $\bm{r}$ in a unit cell.

\begin{figure}[h]
\centering
\includegraphics[scale=0.6]{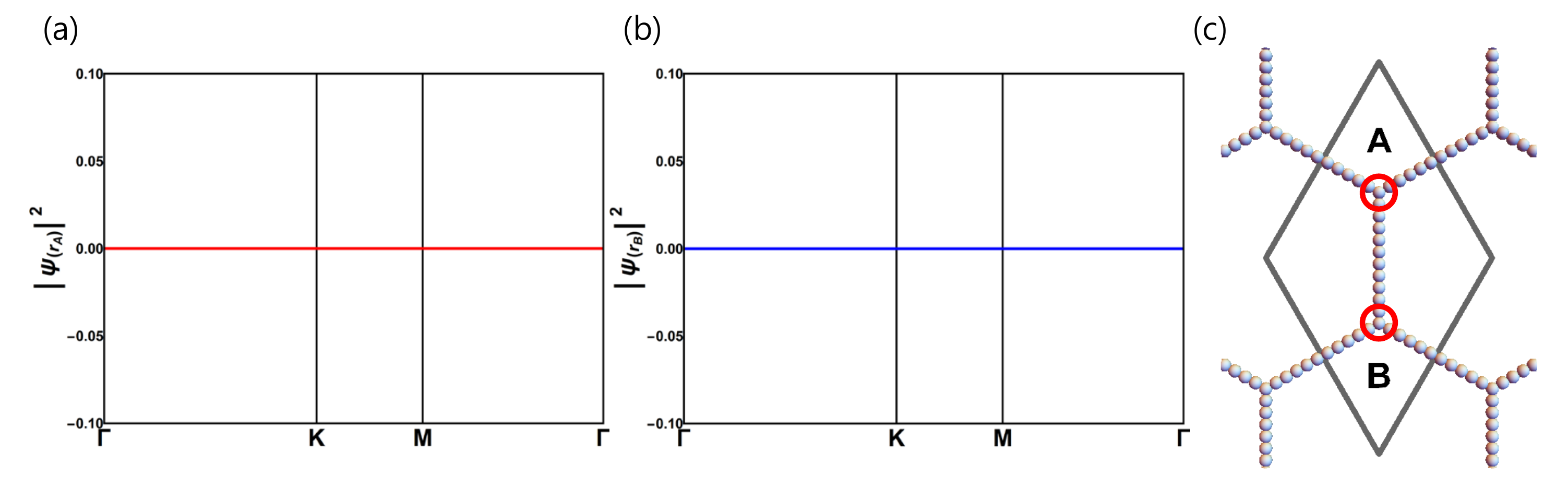}
\caption{The amplitude of Bloch wave function at junction sites in momentum space along the high-symmetry points. (a) is the amplitude at the junction A and (b) is the amplitude at the junction B. We are drawing here only along the high-symmetry cut, but one can confirm that the junction-site wavefunction vanishes everywhere in momentum space. (c) Network and the position of $A$ and $B$ junctions in the unit cell. }
\label{JunctionZero}
\end{figure}

From Fig.\ref{JunctionZero}, we can clearly see that the Bloch function has zeros at the tri-junctions upto the numerical error. This confirms that the flat band states can be generated by the standing waves living in each wires. In fact, we can do better: from the Bloch state, we can even read off the sign structures of the non-dispersing states in the Fig 1 (B) of the main text. The Bloch states have a staggered $\pm 1$ signs around the nodes, which is precisely the same structure as the wavefunction of the main text. 

\begin{figure}[h]
\centering
\includegraphics[scale=0.3]{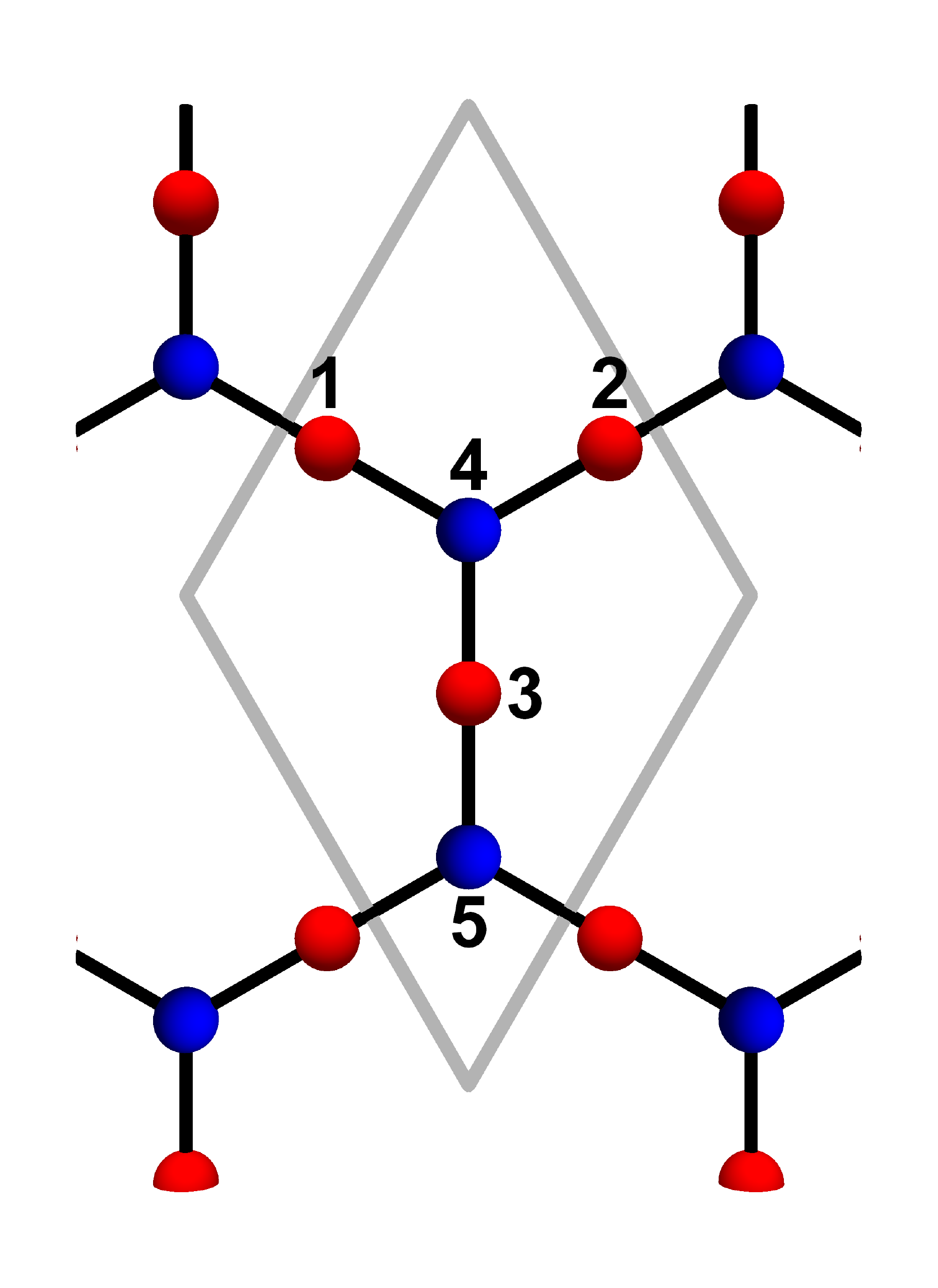}
\caption{Network with a single bridge site.}
\label{Single-Example}
\end{figure}

\subsection{Proof of Existence of A Single Flat Band}
For a few fine-tuned cases, we can in fact prove the existence of a single flat band. The case is that, there are odd number of sites between the tri-junctions and the Hamiltonian has only the nearest neighbor hoppings. For example, for the case with a single site inbetween the tri-junctions Fig. \ref{Single-Example}, we can factorize the Hamiltonian into the following: 

\begin{equation}
H= -\left[ \begin{array}{@{}*{5}{c}@{}}
    0 & 0 & 0 & t_0 & t_0 e^{ika_2} \\
    0 & 0 & 0 & t_0 & t_0 e^{ika_1} \\
    0 & 0 & 0 & t_0 & t_0 \\
    t_0 & t_0 & t_0 & 0 & 0 \\
    t_0 e^{-ika_2} & t_0 e^{-ika_1} & t_0 & 0 & 0 \\
\end{array} \right]
\end{equation}

\begin{figure}[h]
\centering
\includegraphics[scale=0.28]{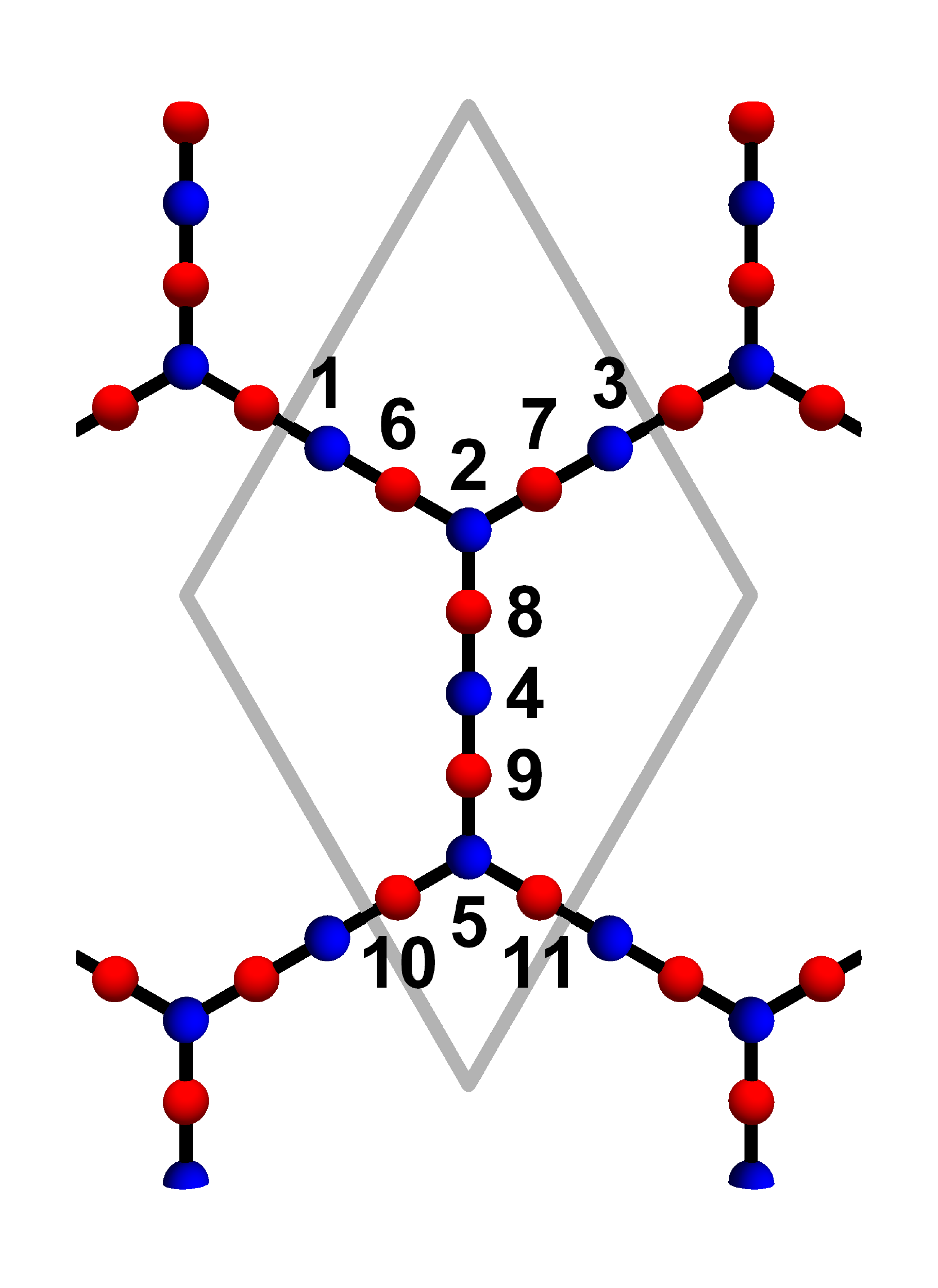}
\includegraphics[scale=0.3]{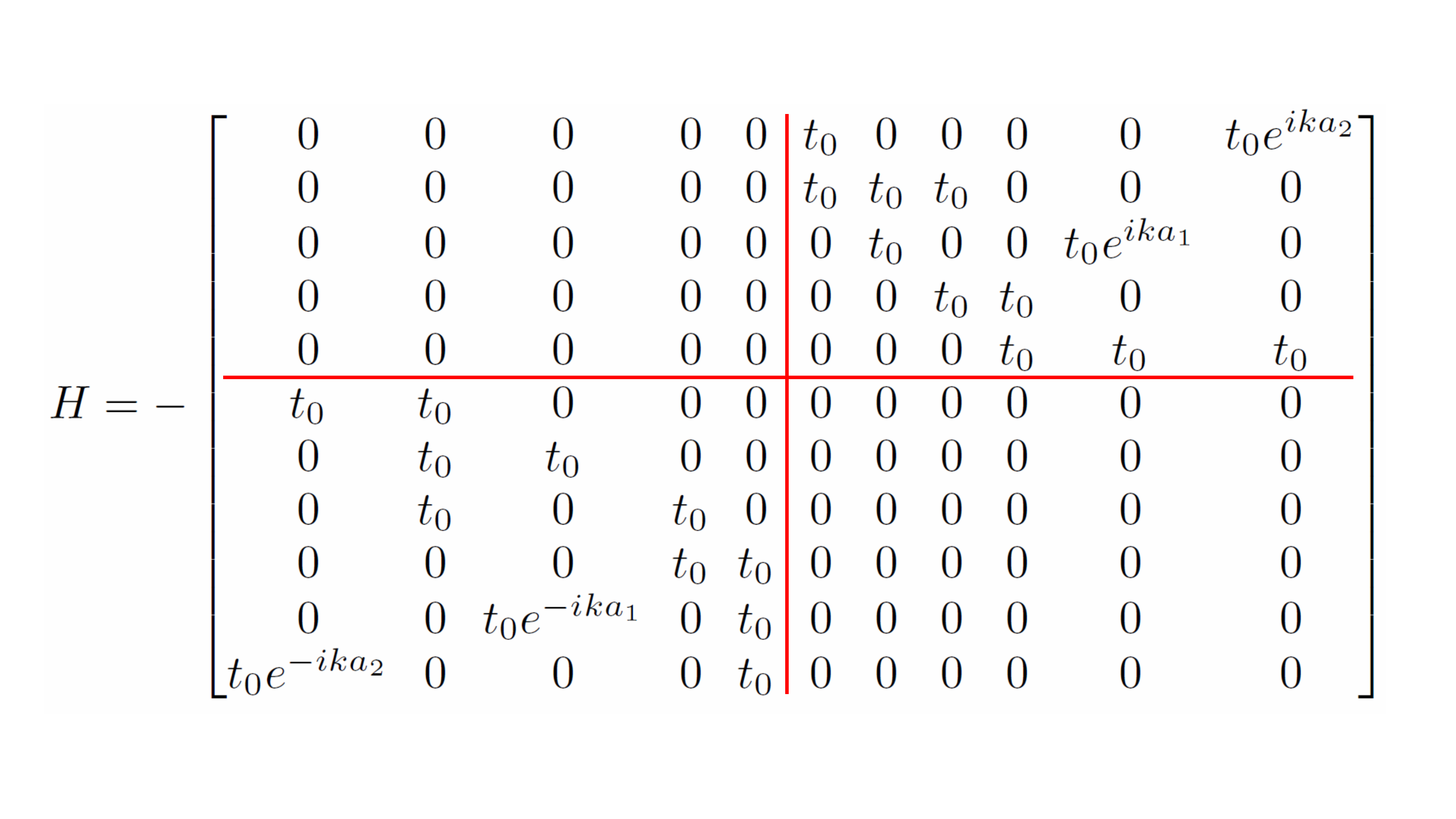}
\caption{(color online) A honeycomb with odd-sites between tri-junctions and only the nearest-neighbor hoppings. The blue sites were indexed from 1 to 5 and the red sited were indexed from 6 to 11. (right) A Hamiltonian matrix followed the indexing method. The red guide line indicates block matrices.}
\label{OffDiagonal}
\end{figure}

Since the Hamiltonian as a matrix has the rank less than its dimension, it must have an eigenvalue $0$ for all the $\bm{q}$. This is the flat band because the corresponding eigenvalue is a constant zero over all the momentum $\bm{q}$. In fact, such factorization can be easily generalized into the cases with the odd number of sites between the tri-junctions. Suppose that $(2m+1)$-sites exist between the tri-junctions, where m is an integer. Hence, there will be $6m+5$ sites per unit cell. We can let the adjacent lattice points be included in different sets A and B since the lattice is bipartite. Assuming that the lattice point at the junction is included in the set A without loss of generality, we find that the set A will have $3m+3$ sites and the set B will have $3m+2$ sites. Now, we index the lattice points in the set A with the integers from $1$ to $3m+3$ and the lattice points in the set B with the integers from $3m+4$ to $6m+5$. Then, since the matrix representation of the tight-binding Hamiltonian is zero for all $(i,j)$ except when i-th and j-th lattice points are adjacent. This makes the Hamiltonian block-off-diagonal, where the blocks are $(3m+3)\times(3m+3)$ and $(3m+2)\times(3m+2)$ sized zero matrices. Hence, the rank of the matrix is lower than the dimension of the matrix by $1$, which signals the emergence of the flat band. Fig.\ref{OffDiagonal} is an example. This honeycomb network has three sites between tri-junctions, and the sites are indexed following the above method. The Hamiltonian, which is $11\times 11$, can be written as the two rectangular off-block-diagonal matrices. 

However, we remark that such factorization is absent (at least we couldn't find it after some trials) for the generic cases with further neighbor hoppings. Hence, this approach cannot be used to prove the existence of many stable flat bands.

\begin{figure}[h]
\centering
\includegraphics[scale=0.3]{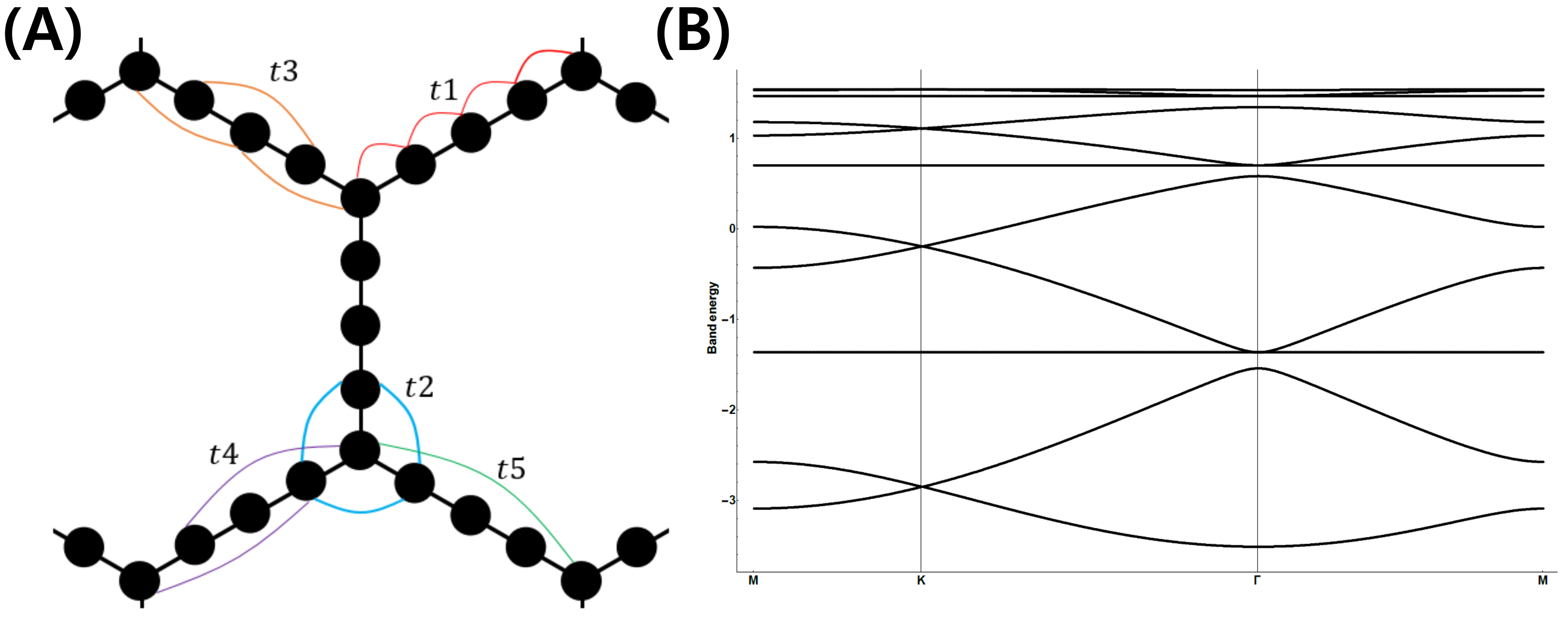}
\caption{Model with intrawire further neighbor terms and its spectrum.}
\label{Intra-model}
\end{figure}

\subsection{Flat Bands with Intrawire Further Neighbor Terms}
Here we demonstrate that the flat bands are intact even in the presence of the intrawire further neighbor terms. See Fig.\ref{Intra-model} for the pictorial representation of our tight-binding model and its spectrum, where we clearly see the intact flat bands.

\section{Flat Bands with Rashba Spin-Orbit Coupling}
\begin{figure*}[h] 
\centering{ \includegraphics[width=17.0 cm]{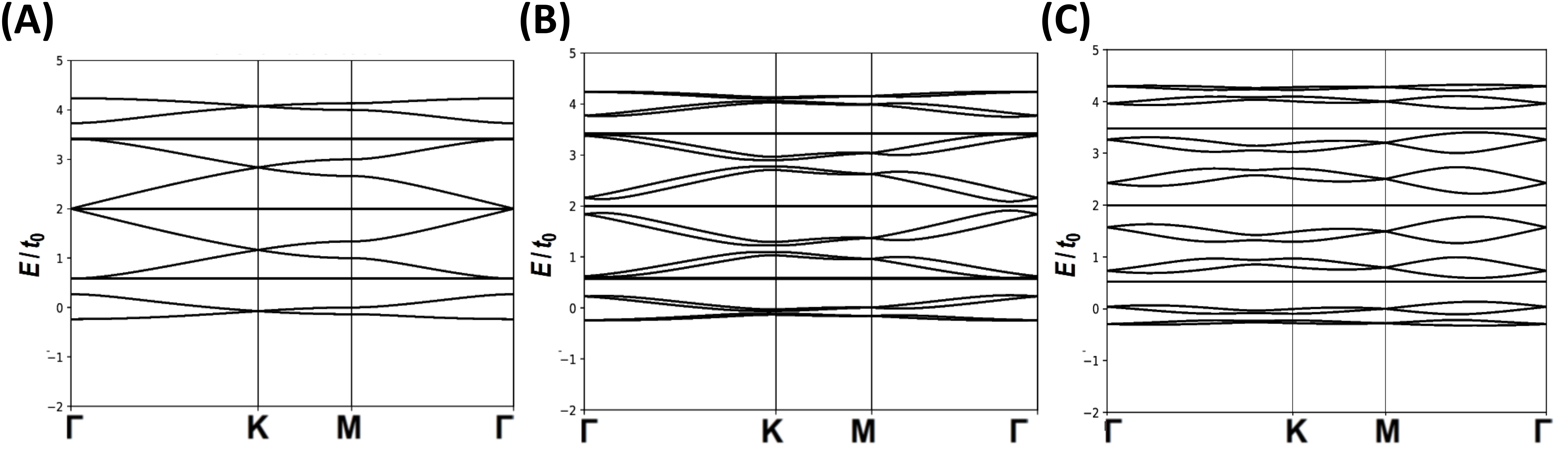} }
\caption{ (A) Band Structure of $\lambda_R =0$. Note the spin-1 Dirac touching at $\Gamma$ at $E=2t_0$. (B) Band Structure of $\lambda_R = 0.1 t$ (C)  Band Structure of $\lambda_R = 0.3 t$. Note that the band touchings are lifted while the flat bands are intact. }\label{SFig-SOC}
\end{figure*}

Here we investigate the effect of the Rashba spin-orbit coupling on the flat bands. Here, the spin-orbit coupling is between the next nearest neighbor sites and we find that this coupling does not disperse the flat bands. 
\begin{align}
H = - t \sum_{\langle \bm{r}, \bm{r}' \rangle} \left( c_{\bm{r}}^{\dagger} c_{\bm{r}'} + h.c.\right) + i \lambda_R \sum_{\bm{r}, \bm{r}'} c_{\bm{r}}^{\dagger} \left[( \vec{s} \times \hat{d}_{\bm{r}, \bm{r}'})\cdot \hat{z}\right] c_{\bm{r}'} + h.c.,
\end{align}
which has the spin-orbit coupling $\lambda_{\bm{R}}$ on top of the tight-binding model Eq.(1) of the main text. Here we set $t_J =0$. We vary $\lambda_R = x \cdot t$ with $x \in [0,1]$. Over the range of $x$, we find the flat bands survive while the band touchings are splitted. See Fig.\ref{SFig-SOC}.

\begin{figure*}[h] 
\centering{ \includegraphics[width=17.0 cm]{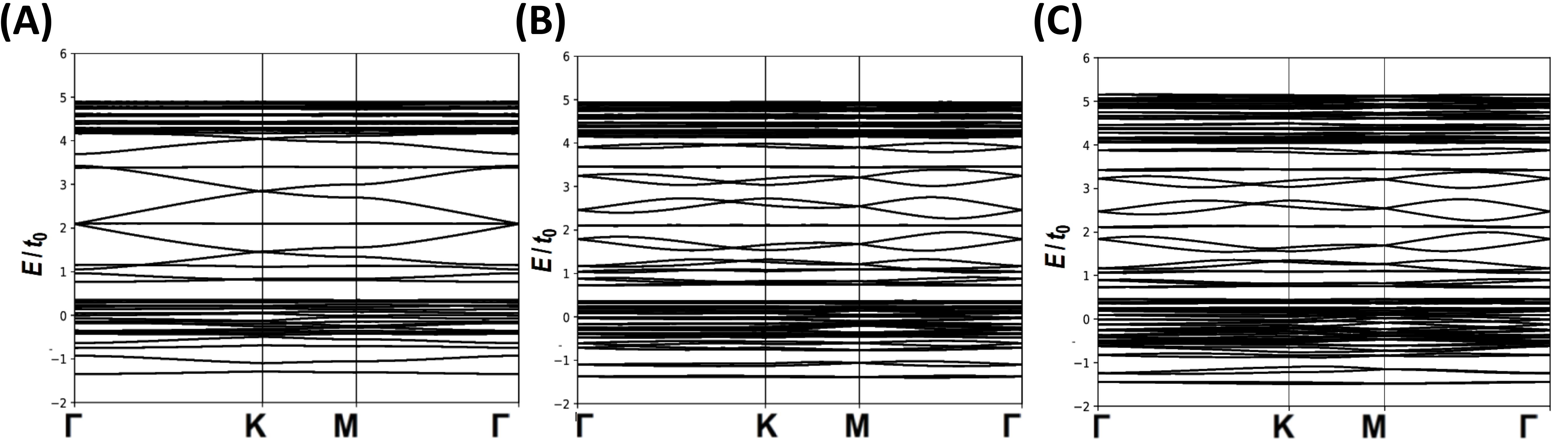} }
\caption{ (A) Band Structure of $\lambda_R =0$. Note the spin-1 Dirac touching at $\Gamma$ at $E=2t_0$. (B) Band Structure of $\lambda_R = 0.3 t$ (for all the second neighbor hoppings) and non-zero spin-symmetric hopping parameters. (C)  Band Structure of different $\lambda_R$ for different pairs of the sites: i.e., two sites between the network, two sites between the domains, and between the domain and the network have the different spin-orbit couplings. Note that the band touchings are lifted while the flat bands are intact. }\label{SFig-SOC2}
\end{figure*}

Next, we include the domain electrons and also all the allowed Rashba spin-orbit couplings, i.e., between the sites within the domain walls, between the site in the domain and the site in networks, and between the sites within the network. This in general will include all the possible symmetry-allowed short-ranged hoppings and spin-orbit couplings. This does not disperse the flat bands much within the bulk gap. See Fig.\ref{SFig-SOC2}. This implies that in TaS${}_2$, even if we include the spin-orbit couplings of Ta atoms, the flat bands will remain intact.

\section{No Stable Flat Bands for Triangular and Square Networks}
\begin{figure*}[h] 
\centering{ \includegraphics[width=17.0 cm]{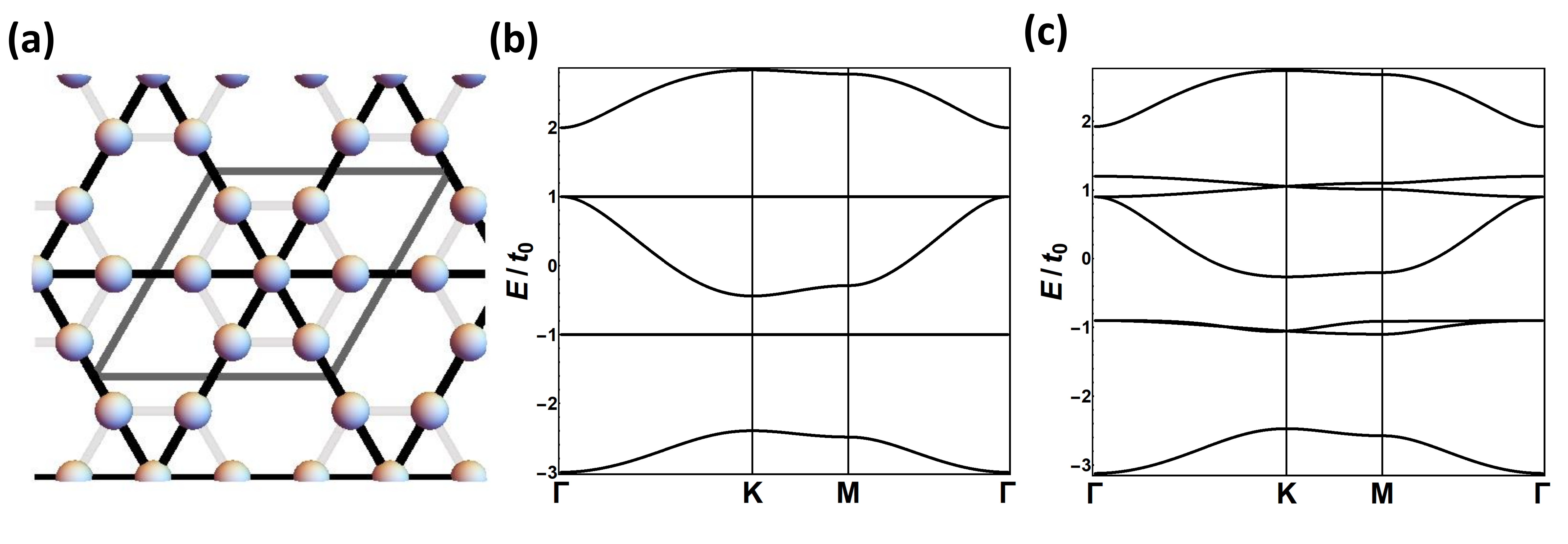} }
\caption{ (a) Triangular Network. (b) Spectrum with only the nearest neighbor hoppings. (c) Spectrum with weak second neighbor hoppings. }\label{Triangle}
\end{figure*}

\begin{figure*}[h] 
\centering{ \includegraphics[width=17.0 cm]{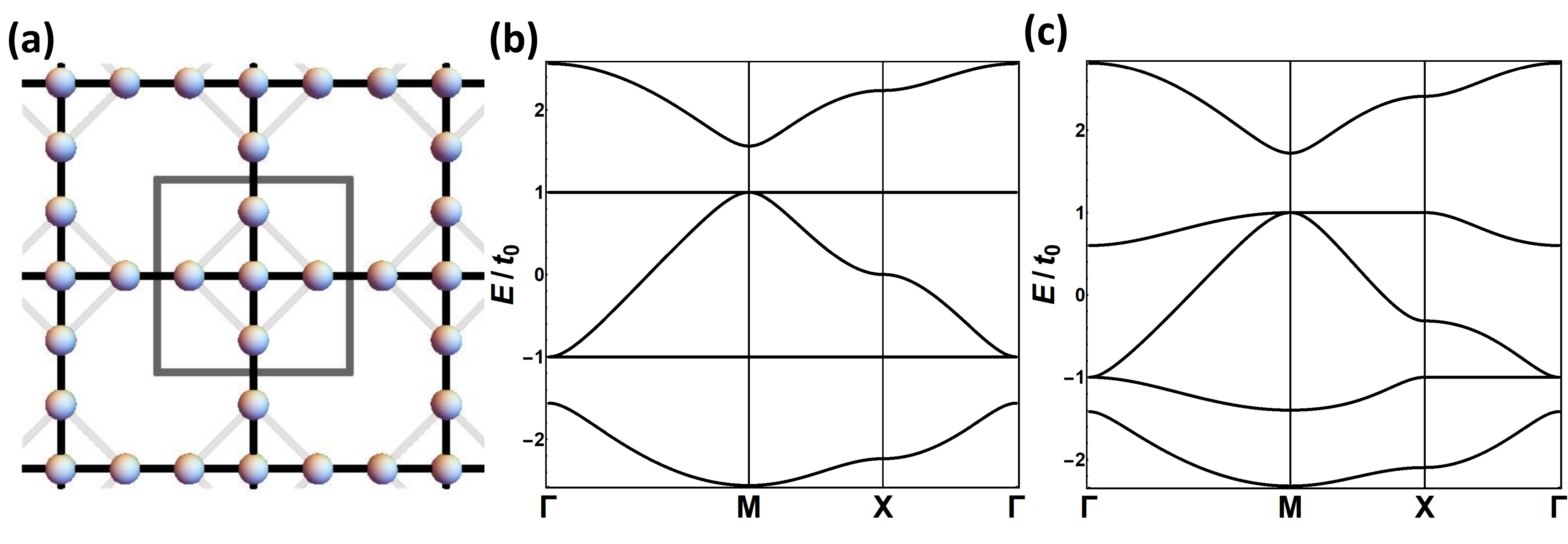} }
\caption{ (a) Sqaure Network. (b) Spectrum with only the nearest neighbor hoppings. (c) Spectrum with weak second neighbor hoppings. }\label{Square}
\end{figure*}

We show that in the triangular or square network Fig. \ref{Triangle} and Fig. \ref{Square}, we do not find any stable flat bands. This is consistent with the fact that there is no delicate destructive interference for these networks. Generically these two networks host the flat bands when there is only the nearest neighbor hoppings. However, they are removed as soon as the second neighbor hoppings are included. 

\section{Effect of Crystal Symmetry Breaking on Flat Bands}

The existence of the flat band heavily relies on the symmetry of the honeycomb network. This directly implies that the flat bands will be dispersive as soon as the protecting crystal symmetries are removed. In this supplemental information, we systematically break the crystal symmetries and show that indeed the flat bands are removed under the breaking of the symmetries. Note that the effect of the $\mathcal{T}$-symmetry breaking  is in the main text, where we found the dispersive Chern bands.

\begin{figure}[h]
\centering
\includegraphics[scale=0.5]{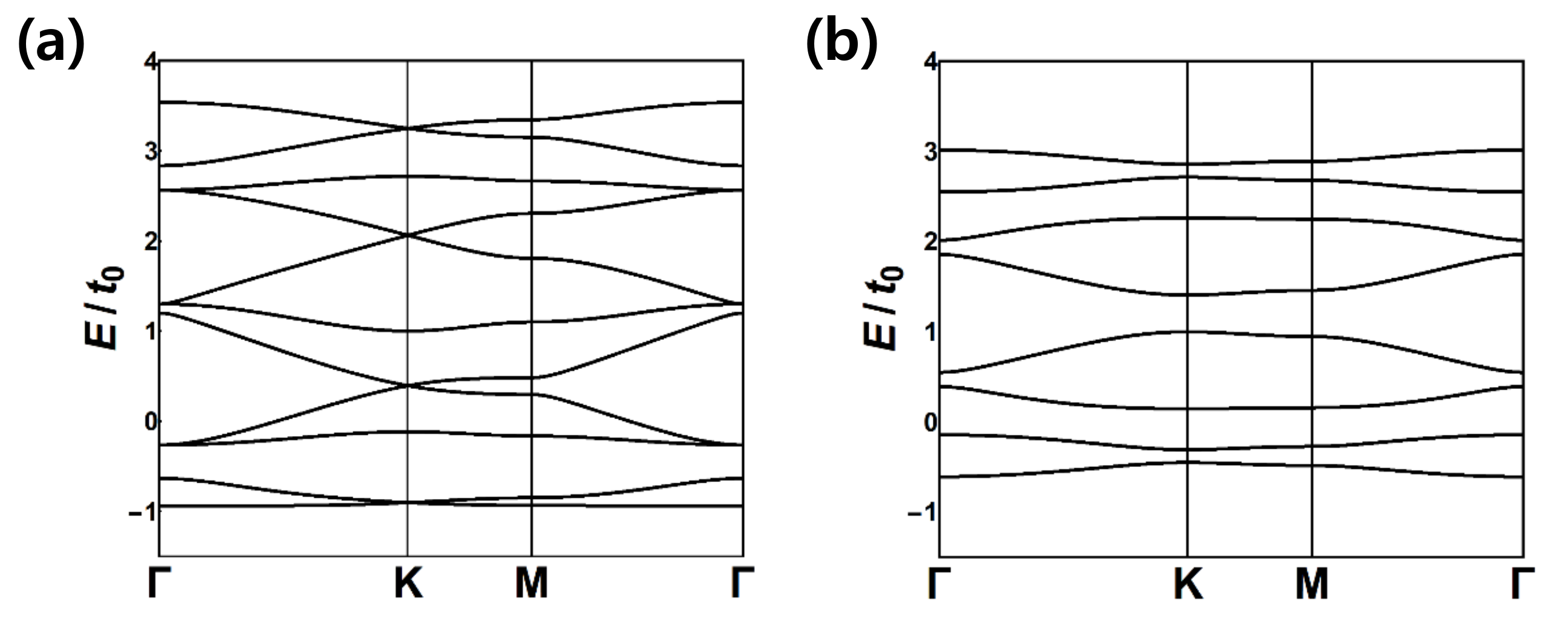}
\caption{Band Structure of symmetry-broken honeycomb network. (a) Typical band structure when the hopping across the domain is added. For this process, there is no destructive interference and so the flat bands become dispersive. (b) Band structure with the broken $\mathcal{T}$-symmetry. In this case, all the band touchings are removed and the flat bands become dispersive. Furthermore, all the bands carry the non-zero Chern numbers. }
\label{SymBrk1}
\end{figure}

\begin{figure}[h]
\centering
\includegraphics[scale=0.5]{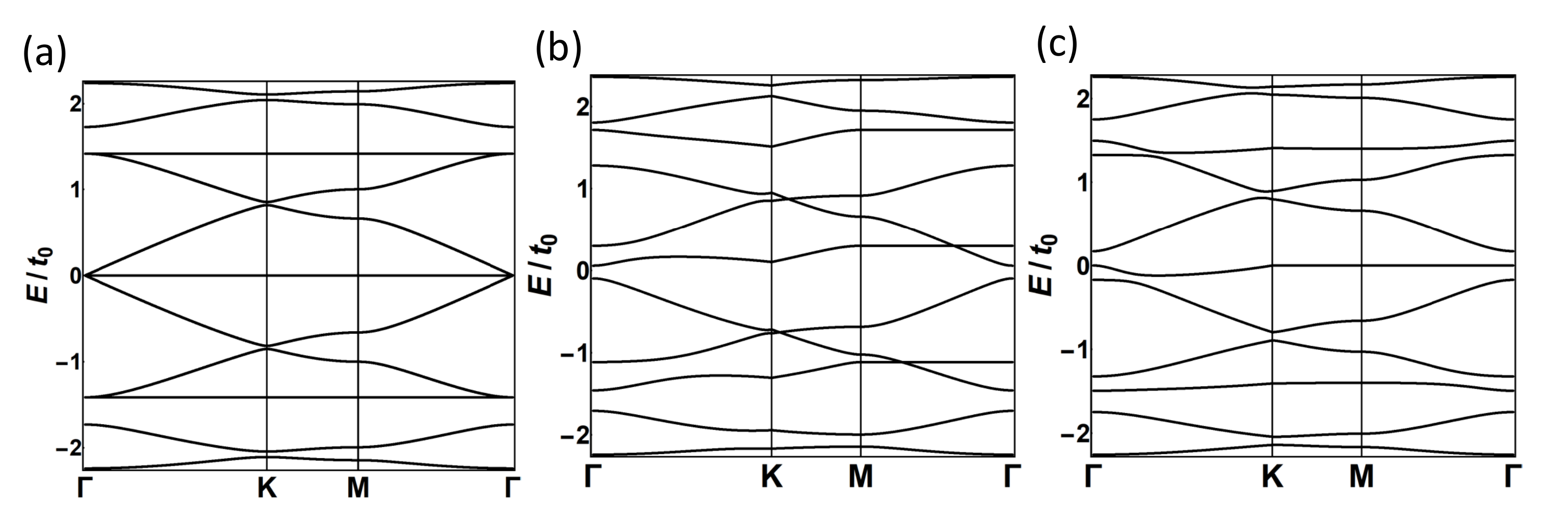}
\caption{Band Structure of symmetry-broken honeycomb network. The symmetries of the corresponding Hamiltonian are (a) $C_3 \times R_x$ symmetry, (b) $C_2$ and $R_x$ symmetry, (c) $C_2$ symmetry. }
\label{SymBrk}
\end{figure}

The band structure of the lower-crystalline symmetries are shown in Fig.\ref{SymBrk1} and Fig.\ref{SymBrk}. The symmetries are broken by extra hoppings or phases in Fig.\ref{SymBrk1} and by on-site potentials $m$ in Fig.\ref{SymBrk}. The matrix representation of the latter Hamiltonians follows the convention of Fig.\ref{OffDiagonal}. Note that the flat bands are removed in all the cases except $C_3 \times R_x$ case where our argument in the main text straightforwardly generalizes. From the previous work, we also have observed that $C_3 \times R_x$ cannot lift the flat bands.\cite{Park-Prep}

\section{Details of BCS Calculations}
Here we present the details of the BCS mean-field calculation. This involves the projection of the bare lattice-scale interactions to the BCS channels and the calculation of the mean-field gap equation and energy. 

\subsection{Coupling Constants}
The microscopic pairing interaction that we introduced phenomenologically in the main text can be generally expressed as
\begin{equation}
H = U\sum_{\bm{R},a} \hat{n}^{2}_{\bm{R};a}
+ V\sum_{\bm{R},\langle a,b \rangle} \hat{n}_{\bm{R};a}\hat{n}_{\bm{R};b}
+ V\sum_{\bm{R},\bm{R}'} \hat{n}_{\bm{R};\text{end}}\hat{n}_{\bm{R}',\text{end'}}
\end{equation}
where U and V are real-valued constants, $\hat{n}_{\bm{R};a}$ is the number operator at position $\bm{R}$ and $a$-th site and $a$, $b$ and ``$\text{end}$" indicate the adjacent sites in real space. We are going to project this interaction terms to the BCS pairing term.

First, we consider the spin degree of freedom then the first term can be expanded.\begin{equation}
U\sum_{\bm{R},a} \hat{n}^{2}_{\bm{R};a} = 
U\sum_{\bm{R},a} (\hat{n}_{\bm{R};a\uparrow}+\hat{n}_{\bm{R};a\downarrow})
(\hat{n}_{\bm{R};a\downarrow}+\hat{n}_{\bm{R};a\uparrow})
\end{equation}
We only select $n_{\uparrow}n_{\downarrow}$ pairs in this expansion since we are interested in the conventional BCS channel. By using the Fourier transformation, the first pairing term becomes
\begin{equation}
\begin{aligned}
= \frac{2U}{N} \sum_{\bm{k}\bm{p}}\sum_{\bm{q}\bm{l}}
\Big[ \delta(\bm{q}+\bm{p}-\bm{l}-\bm{k})
F^{U}(\bm{p},\bm{q};\bm{l},\bm{k})
\psi^{\dagger}_{\bm{p}\uparrow} \psi_{\bm{k}\uparrow}
\psi^{\dagger}_{\bm{q}\downarrow} \psi_{\bm{l}\downarrow}
\Big]
\end{aligned}
\end{equation}
where the form factor is defined as the following. 
\begin{equation}
\begin{aligned}
F^{U}(\bm{p},\bm{q};\bm{l},\bm{k}) = \sum_a u_{\bm{k};a} u^{*}_{\bm{p};a}
u^{*}_{\bm{q};a} u_{\bm{l};a}
\end{aligned}
\end{equation}
Restricting the momentum summation into the pairing channels only, we obtain the BCS channel.
\begin{equation}
\begin{aligned}
U\sum_{R,a} \hat{n}^{2}_{R;a} =& 2U \sum_{\bm{k},\bm{p}} F^{U}_{BCS}(\bm{p},\bm{k})
\psi^{\dagger}_{\bm{p}\uparrow} \psi^{\dagger}_{-\bm{p}\downarrow}
\psi_{-\bm{k}\downarrow} \psi_{\bm{k}\uparrow}\\
F^{U}_{BCS}(\bm{p},\bm{k})=& \sum_a u_{\bm{k};a} u^{*}_{\bm{p};a}
u^{*}_{-\bm{p};a} u_{-\bm{k};a}
\end{aligned}
\end{equation}
Similarly, we calculated the second and the third terms also. 
\begin{equation}
\begin{aligned}
V\sum_{\bm{R},\langle a,b \rangle} \hat{n}_{\bm{R};a}\hat{n}_{\bm{R};b}
+V\sum_{\bm{R},\bm{R}'} \hat{n}_{\bm{R};\text{end}}\hat{n}_{\bm{R}',\text{end}'} 
=& V \sum_{\bm{k},\bm{p}} F^{V}_{BCS}(\bm{p},\bm{k}) 
\psi^{\dagger}_{\bm{p}\uparrow} \psi^{\dagger}_{-\bm{p}\downarrow}
\psi_{-\bm{k}\downarrow} \psi_{\bm{k}\uparrow}\\
F^{V}_{BCS} (\bm{p},\bm{k}) =& \sum_{\langle a,b\rangle} u_{\bm{p};a} u^{*}_{\bm{k};a} u^{*}_{-\bm{p};b} u_{-\bm{k};b} 
\end{aligned}
\end{equation} 
Thus,
\begin{equation}
\begin{aligned}
H =& \sum_{\bm{k},\bm{p}}(2U F^{U}_{BCS}+VF^{V}_{BCS}) 
\psi^{\dagger}_{\bm{p}\uparrow} \psi^{\dagger}_{-\bm{p}\downarrow}
\psi_{-\bm{k}\downarrow} \psi_{\bm{k}\uparrow}\\
=& \sum_{\bm{k},\bm{p}}g(\bm{p},\bm{k}) 
\psi^{\dagger}_{\bm{p}\uparrow} \psi^{\dagger}_{-\bm{p}\downarrow}
\psi_{-\bm{k}\downarrow} \psi_{\bm{k}\uparrow}
\end{aligned}
\end{equation}
Next, we further perform the expansion of $g(\bm{p},\bm{k})$ in terms of the angular momentum sectors, i.e., the s-wave and the d-wave channels. 
\begin{equation}
g(\bm{p},\bm{k}) = \sum_l g_l F_{l}(\bm{k})F^{*}_{l}(\bm{p}) \label{Proj-BCS}
\end{equation}

\subsection{Mean-Field Solution and Energy}
We define the pairing order parameter.
\begin{equation}
\Delta_{l} = g_l \sum_{\bm{k}} F_{l}(\bm{k}) \langle \psi_{l;-\bm{k}\downarrow}\psi_{l;\bm{k}\uparrow} \rangle 
\end{equation}
The form factor $F_l (\bm{k})$ and the $g_l$ is defined as 
\begin{equation}
\begin{aligned}
F_{0}(\bm{k})&=1\\
g_0 &= 2U f^{s}_{U} + V f^{s}_{V}
\end{aligned}
\end{equation}
for the s-wave case, and
\begin{equation}
\begin{aligned}
F_{2}(\bm{k})&=\cos{\sqrt{3}k_x}+e^{\frac{2}{3}\pi i}\cos{\Big( \frac{\sqrt{3}}{2}k_x +\frac{3}{2}k_y\Big) } +e^{-\frac{2}{3}\pi i}\cos{\Big(-\frac{\sqrt{3}}{2}k_x +\frac{3}{2}k_y\Big) }\\
g_2 &= 2U f^{d}_{U} + V f^{d}_{V}
\end{aligned}
\end{equation}
for the d-wave case. Here $\{f^l_U, f^l_V, l =s, d\}$ are the constants depending on a particular choosen flat band. They are obtained from the projection of the BCS channel interactions into the particular pairing channels Eq.\eqref{Proj-BCS}. See the tables below.

\begin{table}[h]
\begin{tabular}{ccccc}
    & $f^{s}_{U}$ & $f^{s}_{V}$ & $f^{d}_{U}$ & $f^{d}_{V}$ \\ \hline
1st & 0.1000      & 0.0958      & 0.0214      & 0.0217      \\ \hline
2nd & 0.1000      & 0.0709      & 0.0214      & 0.0153      \\ \hline
3rd & 0.1000      & 0.0709      & 0.0214      & 0.0153      \\ \hline
4th & 0.1000      & 0.0957      & 0.0214      & 0.0217      \\ \hline
\end{tabular}
\caption{Table of $\{f^l_U, f^l_V, l =s, d\}$ for the system with four sites between the junctions. Note that the band index means the ``n-th" lowest band. }
\end{table}

\begin{table}[h]
\begin{tabular}{ccccc}
    & $f^{s}_{U}$ & $f^{s}_{V}$ & $f^{d}_{U}$ & $f^{d}_{V}$ \\ \hline
1st & 0.0833      & 0.1019      & 0.0178      & 0.0226      \\ \hline
2nd & 0.0833      & 0.0833      & 0.0178      & 0.0178      \\ \hline
3rd & 0.1111      & 0.0000      & 0.0238      & 0.0000      \\ \hline
4th & 0.0833      & 0.0833      & 0.0178      & 0.0178      \\ \hline
5th & 0.0833      & 0.1018      & 0.0178      & 0.0226      \\ \hline
\end{tabular}
\caption{Table of $\{f^l_U, f^l_V, l =s, d\}$ for the system with foive sites between the junctions. Note that the band index means the ``n-th" lowest band. }
\end{table}

With all of these, we can now perform the mean-field theory of Eq. (3) of the main text. Note that $g_l$ should be negative to be attractive.
\begin{equation}
\begin{aligned}
H' &= g_l \sum_{\bm{p}} \Big(F^{*}(\bm{p}) \psi^{\dagger}_{l;\bm{p}\uparrow}\psi^{\dagger}_{l;-\bm{p}\downarrow}\Big)\sum_{\bm{k}} \Big(F(\bm{k})\psi_{l;-\bm{k}\downarrow}\psi_{l;\bm{k}\uparrow}\Big)\\
&=g_l \sum_{\bm{p}} \Big(F^{*}(\bm{p}) \psi^{\dagger}_{l;\bm{p}\uparrow}\psi^{\dagger}_{l;-\bm{p}\downarrow}-F^{*}(\bm{p})\langle \psi^{\dagger}_{l;\bm{p}\uparrow}\psi^{\dagger}_{l;-\bm{p}\downarrow}\rangle+F^{*}(\bm{p})\langle \psi^{\dagger}_{l;\bm{p}\uparrow}\psi^{\dagger}_{l;-\bm{p}\downarrow}\rangle \Big)\\
& \times \sum_{\bm{k}} \Big(F(\bm{k})\psi_{l;-\bm{k}\downarrow}\psi_{l;\bm{k}\uparrow}-F(\bm{k})\langle \psi_{l;-\bm{k}\downarrow}\psi_{l;\bm{k}\uparrow}\rangle+F(\bm{k})\langle\psi_{l;-\bm{k}\downarrow}\psi_{l;\bm{k}\uparrow}\rangle \Big)\\
&=\sum_{\bm{k}} \Big(\Delta_{l} F^{*}(\bm{k}) \psi^{\dagger}_{l;\bm{k}\uparrow}\psi^{\dagger}_{l;-\bm{k}\downarrow} +\Delta^{*}_{l} F(\bm{k})\psi_{l;-\bm{k}\downarrow}\psi_{l;\bm{k}\uparrow}  \Big)   - \frac{|\Delta_{l}|^2}{g_l}
\end{aligned}
\end{equation}
Then, the BCS Hamiltonian is given as the following, ignoring the constant term for a moment.
\begin{equation}
H=\sum_{\bm{k}} \Phi^{\dagger}_{\bm{k}} 
\begin{bmatrix} 
\xi_{\bm{k}} & \Delta_l F^{*}(\bm{k}) \\
\Delta^{*}_l F(\bm{k})  & -\xi_{\bm{k}} 
\end{bmatrix}
\Phi_{\bm{k}}
\end{equation}
where $\Phi_{\bm{k}}$ is the Nambu spinor $(\psi_{\bm{k}\uparrow},\psi^{\dagger}_{-\bm{k}\downarrow})^T$. Here $\xi_{\bm{k}} = \epsilon_{\bm{k}} - \mu$. 

\begin{figure}[h]
\centering
\includegraphics[scale=0.5]{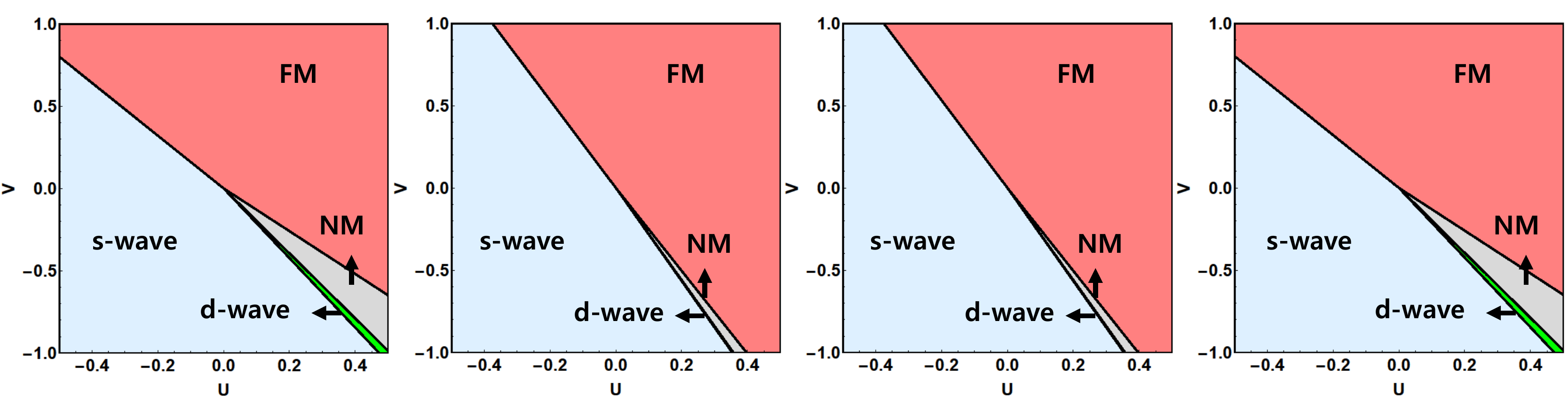}
\caption{Phase diagrams of flat bands of the tight-binding honeycomb network model with 4-bridge sites including the ferromagnetism.(a) the lowest-energy flat band, (b) the second-lowest-energy flat band, (c) the second-highest-energy flat band, (d) the highest-energy flat band. Note that there is always a window for the d-wave superconductors inbetween the s-wave superconductor and the normal metals for $U>0$. }
\label{Phasediagram}
\end{figure}

The $2\times 2$ matrix of the SU(2) group can be expressed with the Pauli matrices.
\begin{equation}
\begin{bmatrix} 
\xi_{\bm{k}} & \Delta_l F^{*}(\bm{k}) \\
\Delta^{*}_l F(\bm{k})  & -\xi_{\bm{k}}
\end{bmatrix} = \bm{n}\cdot \bm{\sigma}
=|\bm{n}| 
\begin{bmatrix} 
\cos{\theta} & e^{-i\phi}\sin{\theta} \\
e^{i\phi}\sin{\theta}  & -\cos{\theta} 
\end{bmatrix}
\end{equation}
We find the unitary matrix $Q$ which diagonalize the Hamiltonian.
\begin{equation}
Q^{\dagger} H Q= 
\begin{bmatrix} 
|\bm{n}| & 0 \\
0  & -|\bm{n}| 
\end{bmatrix}
\end{equation}

\begin{equation}
Q=
\begin{bmatrix} 
\cos{\frac{\theta}{2}} & -e^{-i\phi}\sin{\frac{\theta}{2}} \\
e^{i\phi}\sin{\frac{\theta}{2}}  & \cos{\frac{\theta}{2}} 
\end{bmatrix}
\end{equation}
Then we define states, which are number-conserving, using the eigenvectors.
\begin{equation}
\chi_{\bm{k}} = \begin{bmatrix} 
\alpha_{k\uparrow} \\
\alpha^{\dagger}_{-\bm{k}\downarrow} 
\end{bmatrix}
=Q^{\dagger}\Phi_{\bm{k}}
\end{equation}
\begin{equation}
\begin{aligned}
\alpha_{\bm{k}\uparrow} &= \cos{\frac{\theta}{2}}\psi_{\bm{k}\uparrow}+e^{-i\phi}\sin{\frac{\theta}{2}}\psi^{\dagger}_{-\bm{k}\downarrow}\\
\alpha_{-\bm{k}\downarrow} &= - e^{-i\phi}\sin{\frac{\theta}{2}}\psi^{\dagger}_{\bm{k}\uparrow} + \cos{\frac{\theta}{2}}\psi_{-\bm{k}\downarrow}
\end{aligned}
\end{equation}
The normalized BCS ground state is given as the following.
\begin{equation}
\begin{aligned}
|\Omega_s \rangle &= \Pi_{\bm{k}} \alpha_{k\uparrow} \alpha_{-\bm{k}\downarrow}|\Omega\rangle\\
&= \Pi_{\bm{k}} (\cos{\frac{\theta}{2}}\psi_{\bm{k}\uparrow} +e^{-i\phi}\sin{\frac{\theta}{2}}\psi^{\dagger}_{-\bm{k}\downarrow})
(e^{-i\phi}\sin{\frac{\theta}{2}}\psi^{\dagger}_{\bm{k}\uparrow}-\cos{\frac{\theta}{2}}\psi_{-\bm{k}\downarrow})|\Omega \rangle\\
&\sim \Pi_{\bm{k}} (\cos{\frac{\theta}{2}} -e^{-i\phi}\sin{\frac{\theta}{2}} \psi^{\dagger}_{\bm{k}\uparrow}\psi^{\dagger}_{-\bm{k}\downarrow})
|\Omega \rangle
\end{aligned}
\end{equation}
By definition of the pairing order parameter, we calculate it with the BCS ground state.
\begin{equation}
\begin{aligned}
\Delta_l &= g_l \sum_{\bm{k}} F(\bm{k}) \langle \Omega_s | \psi_{-\bm{k}\downarrow}\psi_{\bm{k}\uparrow} |\Omega_s \rangle\\
&= g_l \sum_{\bm{k}} \Pi_{\bm{p,q}} F(\bm{k}) \langle \Omega | ( \cos{\frac{\theta}{2}} - e^{i\phi}\sin{\frac{\theta}{2}} \psi_{-\bm{p}\downarrow}\psi_{\bm{p}\uparrow}) \psi_{-\bm{k}\downarrow}\psi_{\bm{k}\uparrow} ( \cos{\frac{\theta}{2}} -e^{-i\phi}\sin{\frac{\theta}{2}} \psi^{\dagger}_{\bm{q}\uparrow}\psi^{\dagger}_{-\bm{q}\downarrow} )|\Omega\rangle\\
&= -g_l \sum_{\bm{k}} F(\bm{k}) \sin{\frac{\theta}{2}}\cos{\frac{\theta}{2}}e^{-i\phi}\\
&\simeq -g_l \int d^2 \bm{k} F(\bm{k}) \frac{1}{2} \sqrt{1-\frac{\xi^{2}_{\bm{k}}}{\xi^{2}_{\bm{k}}+|\Delta_l F^{*}(\bm{k})|^2}} \frac{\Delta_l F^{*}(\bm{k})}{|\Delta_l F^{*}(\bm{k})|}
\end{aligned}
\end{equation} 
Assuming that the Fermi energy is exactly at the flat band, we have $\xi_{\bm{k}}=0$.
\begin{equation}
\begin{aligned}
\frac{1}{g_l} 
&\simeq  -\Big(\frac{1}{2} \int d^2 \bm{k} |F(\bm{k})| \Big) \frac{1}{|\Delta_l |}
\end{aligned}
\end{equation} 
Finally, we can get the BCS mean-field energy.
\begin{equation}
\begin{aligned}
\Delta E &= E_{SC}- E_N  \\
&=\Big( -\int d^2 \bm{k} \sqrt{\xi^{2}_{\bm{k}} + \Delta^{2}_{l}|F(\bm{k})|^2 }- \frac{|\Delta_{0}|^2}{g_l}  \Big) - \int d^2 \bm{k} \xi_{\bm{k}}\\
&=\frac{|\Delta_{0}|^2}{g_l} =\Big(\frac{1}{2} \int d^2 \bm{k} |F(\bm{k})| \Big)^2 g_l
\end{aligned}
\end{equation}
As the result of the numerical calculation, the BCS mean-field energies for each possible superconductor type are
\begin{equation}
\Delta E/A^2 \simeq \left\{
  \begin{array}{lr}
     0.25 \times (Vf^{s}_{V}+2Uf^{s}_{u}) &  $(s-wave SC)$\\
     0.2042 \times (Vf^{d}_{V}+2Uf^{d}_{U}) &  $(d-wave SC)$ 
  \end{array}
\right.
\end{equation}
where $A$ is area of the Brillouin zone. With the BCS mean-field energy, we draw the phase diagrams for the tight-binding honeycomb network models. Only the nearest-neighbor hoppings were considered. For the completeness, we also have included the ferromagnetism in Fig.\ref{Phasediagram}. The phase diagrams are drawn at each flat band of the model. We may turn on the next-nearest-neighbor hoppings, but the phase diagrams were slightly changed but they do not induce much difference.

\begin{figure}[h]
\centering
\includegraphics[scale=0.2]{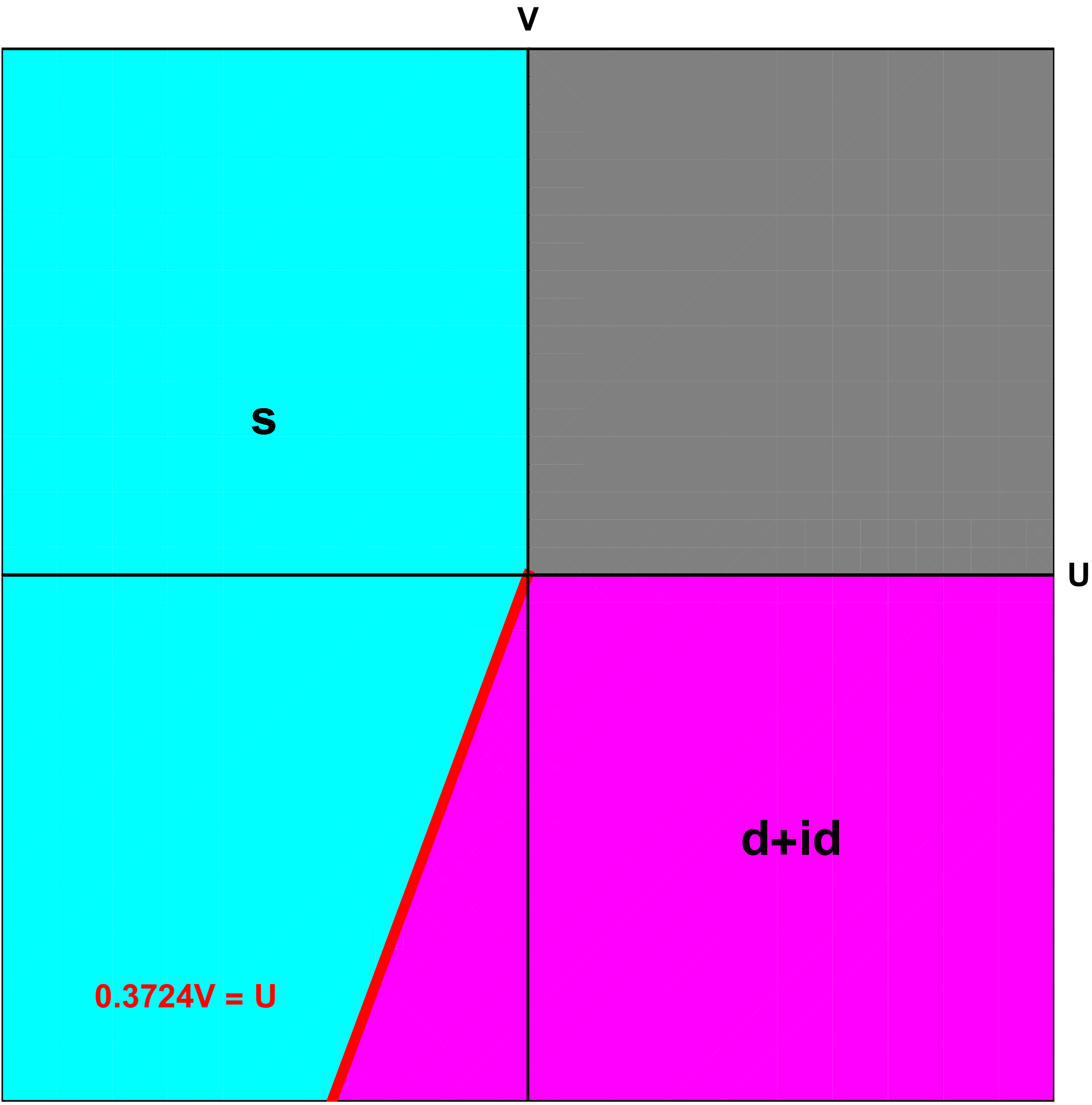}
\caption{Mean-Field Phase diagrams of the square lattice model consist of the interactions U and V away from the half-filling. The black region is for the normal metal.}
\label{Phasediagram-Standard}
\end{figure}

 Note that  the phase space favoring the d-wave pairing is much smaller than that of the s-wave pairing in Fig.\ref{Phasediagram}. This is different from the standard mean-field phase diagram Fig.\ref{Phasediagram-Standard} of the square lattice model consist of the on-site U and nearest-neighbor V. For the square lattice, the system is well away from the half-filling and the Fermi surface is almost circular near the $\Gamma$-point, so that the natural instability is only the pairing. The relative suppression of the d-wave pairing in our flat band model can be understood as following. Note that the electronic Bloch function of the flat band is widely spread in real space. This is very different from the typical cases where the Bloch functions are spread over one or a few sites. Intuitively, electrons in the conventional cases can distinguish clearly the role of U and V because the Bloch state is defined in the same length scale as the scale that distinguishes U and V. Hence, the region for the s-wave and d-wave pairing in the phase space, i.e., (U, V)-plane, can be unambiguously identified. For example, $U>0$ and $V<0$ is the region for the d-wave pairing. On the other hand, in the flat bands, due to the spatial spread of the Bloch states, U and V would be look almost identical to electrons in flat bands. Hence, it is not clear where in the phase diagram the d-wave superconductivity can be favored over the s-wave superconductor. 

 Finally, we woud like to mention a few words on the effect of the phonons in the mean-field phase diagram. As explained in the main text, the soft phonons are likely hardened in the NC-CDW state and hence the phonon-electron couplings are weak. The phonon-electron coupling can be written in the form of 
\begin{align}
H_{ph-el} = g_{p} \int d\bm{r} ~\Phi(\bm{r}) \sum_{\sigma} \psi^{\dagger}_{\sigma}(\bm{r}) \psi_\sigma(\bm{r}). 
\end{align}
Here $\Phi(\bm{r})$ is the phonon field. When the phonon is integrated out, we find the attraction in the BCS channel: 
\begin{align}
H_{eff} = - \frac{g_{p}^{2}}{\omega_{D}} \int d\bm{k} d\bm{k}'  \psi^{\dagger}_{\uparrow}(\bm{k})\psi^{\dagger}_{\downarrow}(-\bm{k}) \cdot \psi_{\downarrow}(\bm{k}')\psi_{\uparrow}(-\bm{k}'), 
\end{align} 
where $\omega_{D}$ is the Debye frequency. At the level of the mean-field theory, he term has almost the same effect as the negative U. Hence, the addition of the electron-phonon coupling explicitly into the model is equivalent to the shift of U toward negative. Thus, we can first conclude that the phonon-electron coupling is included in our effective (U,V)-interactions. Second, obviously, this phonon-electron favors the s-wave pairing and enlarges the area covered by the s-wave pairing. This helps to stabilize the s-wave pairing. Hence, our main conclusion of the mean-field diagram, i.e., the robust superconductivity seen in the experiment is likely the s-wave pairing because the dominant instability of the flat band is the s-wave superconductivity, is intact from the inclusion of this phonon-electron coupling.

\subsection{BdG Fermion Spectrum}
\begin{figure}[h]
\centering
\includegraphics[scale=0.5]{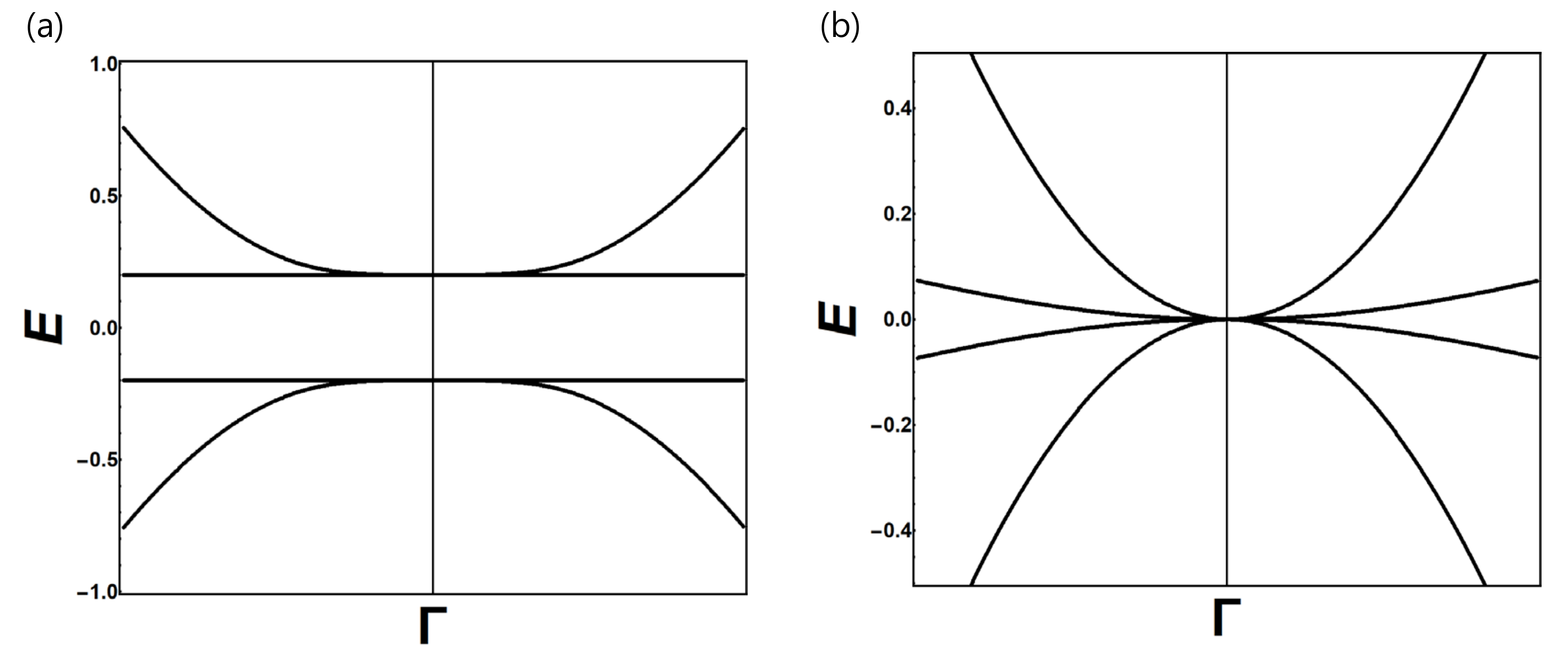}
\caption{The BdG spectrum for (a) the $s$-wave superconductor and (b) the $d$-wave superconductor near the $\Gamma$-point.}
\label{BdG_Spectrum}
\end{figure}

We draw the BdG fermion spectrum.
\begin{equation}
\begin{aligned}
\xi_{\bm{k}} =& \epsilon_{0} \tau^{0} +\sum^{3}_{i=1} \epsilon_{i} \tau^{i}\\ 
=& \frac{1}{2m} \left[ \begin{array}{@{}*{2}{c}@{}}
    (k_x +k_y)^2 &k^{2}_{x}-k^{2}_{y} \\
    k^{2}_{x}-k^{2}_{y} & (k_x -k_y)^2
\end{array} \right],
\end{aligned}
\end{equation} 
which has the flat band with the quadratic band touching at $\bm{k} =0$. With the BCS pairing interaction, the fermion spectrum becomes the following.
\begin{equation}
E(\bm{k}) = \pm \sqrt{(k^2 \pm k^2)^2/4m^2 + |\Delta_{l}F(\bm{k})|^2},
\end{equation}
hence there are four bands at the low-energy limit. The BdG fermion spectra of the s-wave and the d-wave are plotted in Fig.\ref{BdG_Spectrum} near the $\Gamma$-point. The $(d\pm id)$-wave SC is not fully gapped and exhibits the doubled quadratic band touching. Since this state is gapless, it does not support any topologically-protected edge mode. Because the quadratic band touching is marginally unstable\cite{PhysRevLett.103.046811} against the short-ranged repulsive interactions toward the chiral or nematic states, there will be a successive transition at the temperatures below the SC $T_c$. Once the system undergoes a transition to the chiral state, it becomes truly a topological superconductor with the quantized thermal Hall response. This state potentially has an interesting quantum critical behavior, which we leave for the future study. 

\section{Corner States at Tri-Junctions}
In this supplemental information, we provide a few concrete realizations of the zero-dimensional states localized at the tri-junctions (or nodes) of the honeycomb network. We concentrate on the half-filled per wire case, in which the leading insulating instability is the period-2 charge-density waves. 

The zero-dimensional states is the soliton of the charge-denstiy waves, and they carry the fractional quantum numbers, e.g., the electric charge. Here we assume SU(2) symmetry and so consider effectively spinless fermions. Now, imagine that the tri-junction induces the frustrations for the charge-density wave order parameters between the neighboring links of the honeycomb plaquette. Due to the commensurability of the filling, the only allowed frustration is the $\pi$-phase shift between the neighboring links and is trapped at the junction. Once such mode is trapped at the junction, it cannot move around when the crystalline symmetries are protected. [Here we do not allow the Hilbert space to be changed when we consider the symmetric deformations of the model.] Then, we find that the 2d domains are insulating and its first-order boundaries are also insulating, but only its second-order boundary, i.e., the corner, has ``in-gap" modes, which can be protected by the crystal symmetries. This is very parallel with the corner modes in the higher-order topology, (or more precisely, obstructed atomic insulator with the non-trivial nested Wilson loop topology).\cite{Benalcazar61, HOTI_3, HOTI_4, kang2018many} Note that, in the reference\cite{PhysRevB.99.155102}, the junctions of gapped wires are considered in an entirely different context, and the emergent corner states are discovered. Our finding is consistent with theirs. 

It is well-known that, in 1d, a soliton of the period-2 charge density waves is the same as the boundary mode of the Su-Schieffer-Heeger model. In fact, our construction here is essentially the charge-density wave verions of the higher-order topology. In this paper, we will not attempt to present the full theory and classification. Instead, we will present only the minimal contents, and the precise connection and classification of the ``higher-order topological" domain wall states will be reported elsewhere. 

\subsection{Tight-Binding Models} 

\begin{figure}[h]
\centering
\includegraphics[scale=0.55]{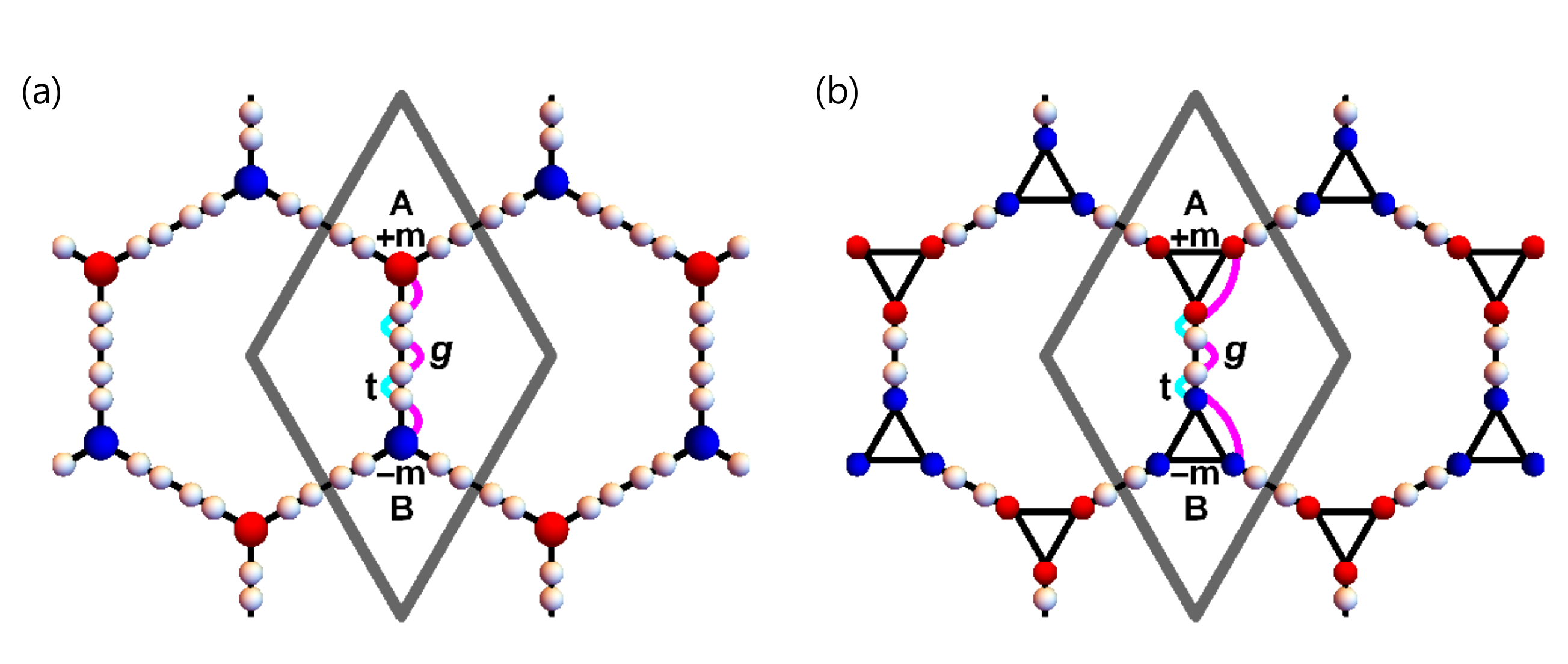}
\caption{(a) An unit cell(gray line) of the model is plotted in real space. Short and long atomic distances on the figure indicates hopping integrals $t$ and $g$ respectively. Infinitesimal on-site energy, $+m$ and $-m$, were given at the site A and the site B. (b) The localized charge at each site, A(blue) and B(red), were plotted by the hopping integral's ratio.  (c) A Hamiltonian matrix of the model in the momentum space.}
\label{CornerState}
\end{figure}

Two examples of tight binding models exhibiting the localized corner state is shown in Fig.\ref{CornerState}. Because of the period-2 modulation, we modulate $t$ and $g$ in each link. Infinitesimal on-site energies, $\pm m$ ,were given at each junctions to break a symmetry weakly (this is commonly done in the investigation of the corner charge in higher-order topology and polarization chain).\cite{Benalcazar61, HOTI_3, HOTI_4, kang2018many} They are explicitly written in Eq.\ref{Corner_Matrix} and Eq.\ref{No_Corner_Matrix}. Then, we calculated the localized charges at each junction at half-filling. The charges at A(Red) and B(Blue) site showed opposite signs but their amplitude was nearly $0.5$ when $g$ is smaller enough than $t$ (i.e., small correlation length limit). The better localization of electric charge is expected when more sites are assumed on the wires as the Su-Schrieffer-Higger model does.        

\begin{equation}
H= -\left[ \begin{array}{@{}*{14}{c}@{}}
    0 &0 &t &0 &0 &0 &0 &0 &0 &0 &0 &0 &0 &ge^{ika_2} \\
    0 &0 &0 &t &0 &0 &0 &0 &0 &0 &0 &0 &ge^{ika_1} &0 \\
    t &0 &0 &0 &g &0 &0 &0 &0 &0 &0 &0 &0 &0 \\
    0 &t &0 &0 &g &0 &0 &0 &0 &0 &0 &0 &0 &0 \\
    0 &0 &g &g &+m &g &0 &0 &0 &0 &0 &0 &0 &0 \\
    0 &0 &0 &0 &g &0 &t &0 &0 &0 &0 &0 &0 &0 \\
    0 &0 &0 &0 &0 &t &0 &g &0 &0 &0 &0 &0 &0 \\
    0 &0 &0 &0 &0 &0 &g &0 &t &0 &0 &0 &0 &0 \\
    0 &0 &0 &0 &0 &0 &0 &t &0 &g &0 &0 &0 &0 \\
    0 &0 &0 &0 &0 &0 &0 &0 &g &-m &g &g &0 &0 \\
    0 &0 &0 &0 &0 &0 &0 &0 &0 &g &0 &0 &t &0 \\
    0 &0 &0 &0 &0 &0 &0 &0 &0 &g &0 &0 &0 &t \\
    0 &ge^{-ika_1} &0 &0 &0 &0 &0 &0 &0 &0 &t &0 &0 &0 \\
    ge^{-ika_2} &0 &0 &0 &0 &0 &0 &0 &0 &0 &0 &t &0 &0 
\end{array} \right]
\label{Corner_Matrix}
\end{equation}

\begin{equation}
H= -\left[ \begin{array}{@{}*{12}{c}@{}}
    0 &0 &t &0 &0 &0 &0 &0 &0 &0 &0 &ge^{ika_2}\\
    0 &0 &0 &t &0 &0 &0 &0 &0 &0 &ge^{ika_1} &0\\
    t &0 &+m &g &g &0 &0 &0 &0 &0 &0 &0\\
    0 &t &g &+m &g &0 &0 &0 &0 &0 &0 &0\\
    0 &0 &g &g &+m &t &0 &0 &0 &0 &0 &0\\
    0 &0 &0 &0 &t &0 &g &0 &0 &0 &0 &0\\
    0 &0 &0 &0 &0 &g &0 &t &0 &0 &0 &0\\
    0 &0 &0 &0 &0 &0 &t &-m &g &g &0 &0\\
    0 &0 &0 &0 &0 &0 &0 &g &-m &g &t &0\\
    0 &0 &0 &0 &0 &0 &0 &g &g &-m &0 &t\\
    0 &ge^{-ika_1} &0 &0 &0 &0 &0 &0 &t &0 &0 &0\\
    ge^{-ika_2} &0 &0 &0 &0 &0 &0 &0 &0 &t &0 &0
\end{array} \right]
\label{No_Corner_Matrix}
\end{equation}

These models are designed in the way that they trap the odd number of solitons at the tri-junction. Once the solitons are trapped, obviously they can be protected by the crystalline symmetries.  

\section{Corner States at Junction of Two domain walls}
\begin{figure}[h]
\centering
\includegraphics[scale=0.5]{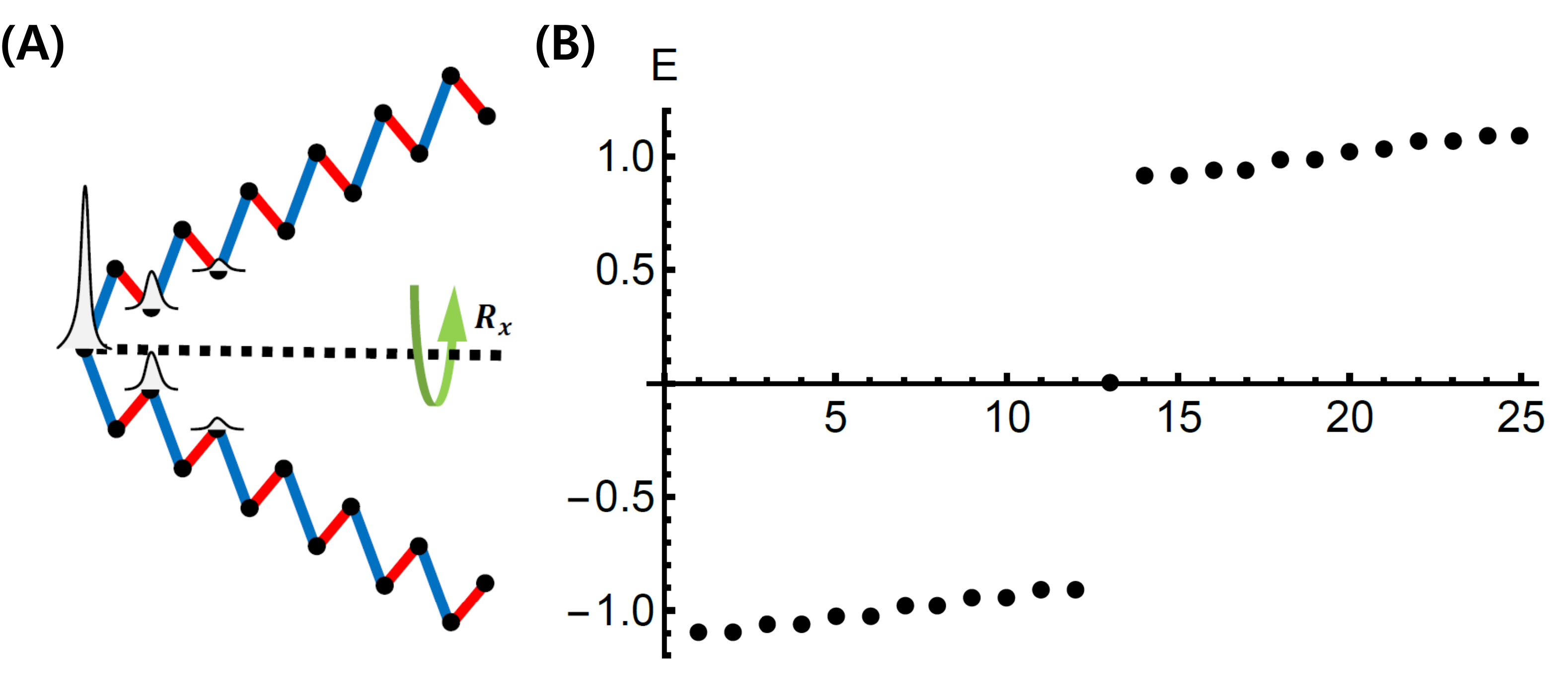}
\caption{(A) Junction of two SSH chain. Here the red bond represents the stronger hopping, $t$ and the blue bond represents the weaker hopping $g$. The peaks represent the probability distribution of the single in-gap mode in the spectrum (B). We have exaggerated the probability amplitude of the in-gap state at the sites away from the junction here, e.g., more than 99\% of the wavefunction probability is localized at the junction for $g/t =  0.1$. (B) Spectrum of the finite size calculation with $g/t =  0.1$. Note the emergence of the single in-gap mode. }
\label{DWJ}
\end{figure}

\begin{figure}[h]
\centering
\includegraphics[scale=0.5]{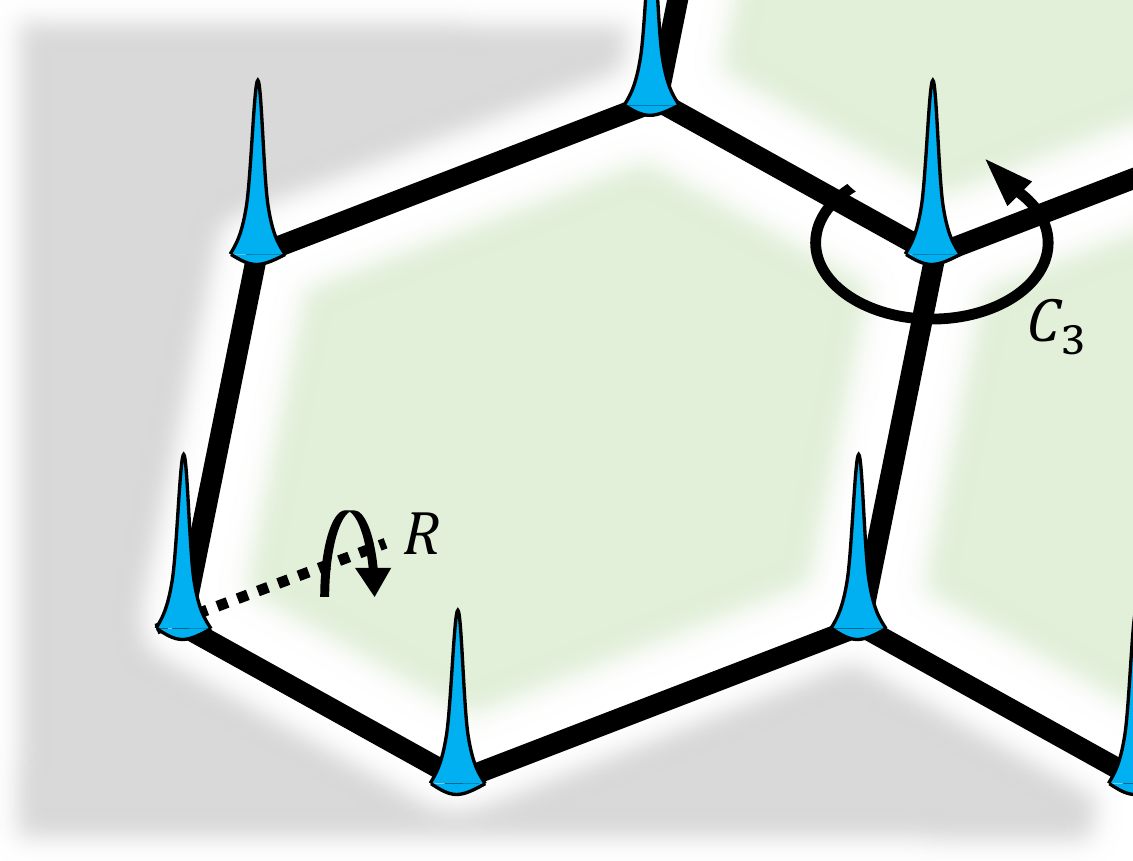}
\caption{Density distributions of in-gap modes which are protected by crystalline symmetries. On the boundary, the reflection symmetry protects the localized in-gap mode, and the $C_3$ crystalline symmetry protects the mode at the junction. Because the system is gapped and thus the correlation length is finite, it is enough to have the crystal symmetries locally near the corners. }
\label{DWJJ}
\end{figure}

Here we present the detailed tight-binding model for the corner state at the junction of the two gapped domain walls. Note that this type of the junction has been observed in the STM experiment of C-CDW 1T-TaS${}_2$\cite{cho2017correlated, PhysRevLett.122.036802}, where the domain wall of C-CDW 1T-TaS${}_2$ is effectively equivalent to the famous Su-Schieffer-Heeger (SSH) chain. Motivated from the experiment, here we construct a minimal model supporting a reflection-symmetry-protected corner state, which is consist of the two SSH chains. See Fig.\ref{DWJ}. Note that the model localizes a single state at the junction, which is protected by the crystalline $R_x$ symmetry. One can view this system as one corner of the edge of the network, when the network has an edge. See the edge of the network in Fig.\ref{DWJJ}. The emergence of the crystal-symmetry-protected 0d state at the corner of the system is the hallmark of the 2d higher-order topology.\cite{Benalcazar61, HOTI_3, HOTI_4, kang2018many}
 
Combining the result in this section with the in-gap modes at the junctions of the network, i.e., Fig. 2 of the main text, we find the local density of state plot, schematically represented in Fig. \ref{DWJJ}, which clearly demonstrate the emergence of crystal-symmetry-protected in-gap modes.

\section{Thermodynamics of Flat Bands}
Here we compute the thermodynamic quantities of the flat bands. Since the contribution from the band touching will be negligible, we will take a completely flat band without the band touching here to compute the quantities. 

\subsection{Specific Heat}

\begin{figure}[h]
\centering
\includegraphics[scale=0.5]{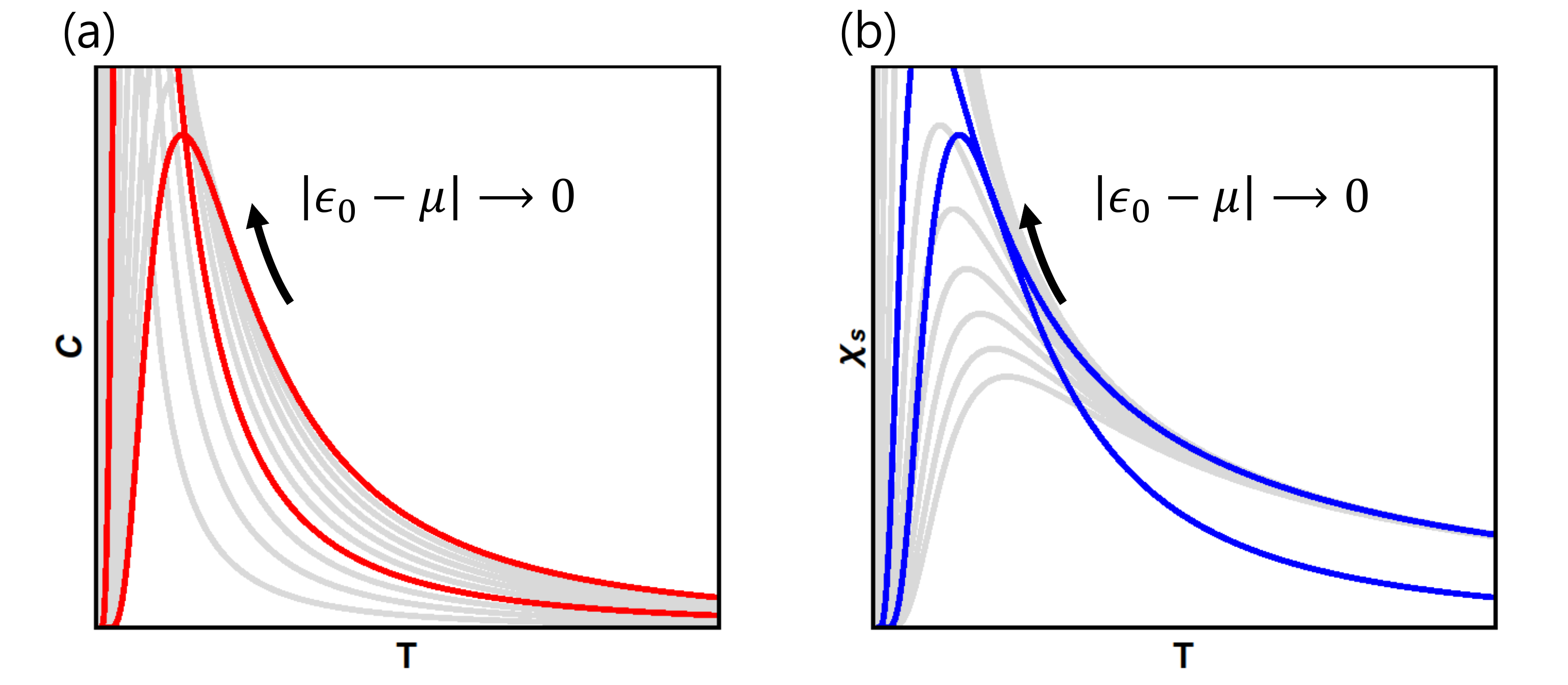}
\caption{Plot of the specific heat and the spin susceptibility of the flat bands.}
\label{Fig12}
\end{figure}

The total energy U is given as the following. 
\begin{align}
U= \int^{\infty}_{0} \epsilon &D(\epsilon) f(\epsilon,T) d\epsilon
\end{align}
where $f(\epsilon,T)$ is the Fermi-Dirac distribution. By the definition of the specific heat,
\begin{equation}
\begin{aligned}
C=&\frac{\partial U}{\partial T}
= \int^{\infty}_{0} \epsilon D(\epsilon) \frac{\partial}{\partial T}\Big(\frac{1}{e^{(\epsilon-\epsilon_F)/k_B T}+1}\Big) d\epsilon
\end{aligned}
\end{equation}
The number of state at the flat band is $N\delta(\epsilon-\epsilon_0)$ (with $N$ being the number of states at the flat band) where $\epsilon_0$ is the energy of the flat band. 
\begin{equation}
\begin{aligned}
C=& \frac{N\epsilon_0 (\epsilon_0 -\epsilon_F)}{k_B T^2} \frac{e^{(\epsilon_0-\epsilon_F)/k_B T}}{(e^{(\epsilon_0-\epsilon_F)/k_B T}+1)^2}
\end{aligned}
\end{equation}
Therefore, near the flat band, $|\epsilon_0-\epsilon_F| \ll k_B T$, we find 
\begin{equation}
C \sim \left\{
  \begin{array}{lr}
    0 &  (\epsilon_0=\epsilon_F)\\
    \frac{1}{T^2} &  (\epsilon_0\neq\epsilon_F) 
  \end{array}
\right.
\end{equation}
Note that, when the Fermi level is exactly at the flat band, the specific heat is always zero. We plot the specific heat in Fig.\ref{Fig12}. At the low-temperature limit, the specific heat is suppressed by $\sim \exp (-|\epsilon_0-\epsilon_F|/k_B T)$. 

\subsection{Spin Susceptibility}

We consider a spin-1/2 system where the both spin species have the same flat band energy at $E_0$. To compute the spin susceptbility, we apply the magnetic field $h$ parallel to the z-axis.  
\begin{equation}
H'=-h \sigma^z
\end{equation}
The flat bands are split into two flat bands at $(E_0+h)$ for spin-down and $(E_0-h)$ for spin-up so the density of states is given as the following.
\[  \left\{
  \begin{array}{lr}
    D_{\uparrow}(\epsilon) = \frac{N}{2}\delta(\epsilon-(\epsilon-h)) \\
    D_{\downarrow}(\epsilon) = \frac{N}{2}\delta(\epsilon-(\epsilon+h))
  \end{array}
\right.
\] 
The number of occupied electrons for each flat band is
\begin{equation}
N_\sigma = \int D_{\sigma}(\epsilon)f(\epsilon,\mu)
\end{equation}
where $\sigma=\uparrow,\downarrow$ and $f(\epsilon,\mu)$ is the Fermi-Dirac distribution.

Magnetization is proportional to the difference of number of the spin-up and spin-down electrons. 
\begin{equation}
\begin{aligned}
m=\frac{1}{2}(N_{\uparrow} -N_{\downarrow}) =&\frac{N}{4} \int^{\infty}_{0} \Big( \delta(\epsilon-(\epsilon_0 -h))-\delta(\epsilon-(\epsilon_0 +h)) \Big) \frac{1}{e^{(\epsilon-\mu)/k_B T}+1} d\epsilon\\
=& \frac{N}{4} \Big(\frac{1}{e^{(\epsilon_0 -h-\mu)/k_B T}+1} -\frac{1}{e^{(\epsilon_0 +h-\mu)/k_B T}+1}  \Big)
\end{aligned}
\end{equation}  

By definition of the spin susceptibility,
\begin{equation}
\begin{aligned}
\chi =& \frac{\partial m}{\partial h} \Big|_{h=0}\\
 =& \frac{N}{2k_B T} 
\frac{e^{(\epsilon_0-\mu)/k_B T}}{(e^{(\epsilon_0-\mu)/k_B T}+1)^2} 
\end{aligned}
\end{equation}

The spin susceptibility near the flat band shows a similar tendency with the specific heat as indicated in Fig.\ref{Fig12}. The spin susceptibility is zero at the zero-temperature. As the fermi energy approach to the flat band energy, the spin susceptibility diverges at the zero-temperature.

\section{Tight-Binding Model for Network in NC-CDW 1T-TaS${}_2$}
Here we introduce some relevant information of the tight-binding model and its comparison with DFT+U calculation for the realistic domain wall network of NC-CDW 1T-TaS${}_2$. The details of the tight-binding model and DFT+U calculation on the domain walls will be reported elsewhere in the near future. 

\begin{figure}[h]
\centering
\includegraphics[scale=0.5]{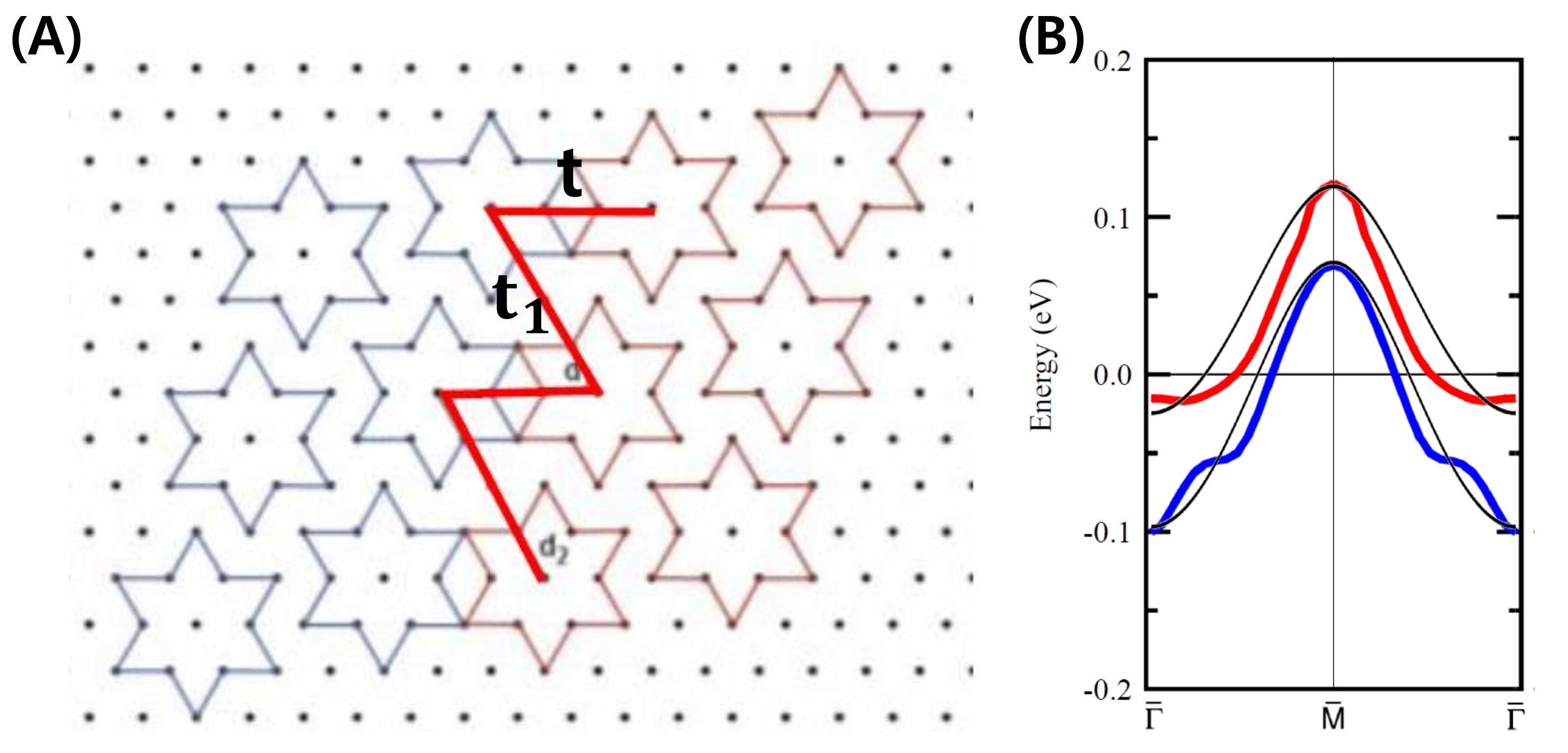}
\caption{(A) Schematic picture of the domain wall. Centers of David stars connected by thick red lines are the domain wall sites, which enter into the effective zig-zag tight-binding model. (B) Comparison of DFT+U calculation and tight-binding calculation. Here the thick red and blue are the domain wall bands near the Fermi level obtained by DFT calculation, and the black lines are the results of the tight-binding model.}
\label{DW-1}
\end{figure}

Here we perform the DFT simulation of a single domain wall surrounded by several David stars (see, e.g., our previous paper\cite{Park-Prep} for details), and then we extract the relevant tight-binding parameters by fitting the bands near the Fermi level obtained from DFT with a tight-binding model\cite{PhysRevLett.122.036802} Fig.\ref{DW-1} (A). This is because a DFT calculation of a single unit cell of the network superstructure, whose size is roughly $\sim$O($10$) nm, is computationally costly and so the DFT calculation cannot be directly done for the superstructure. When we perform fitting of the tight-binding model with the DFT result, we scaled the tight-binding parameters as $1/d^5$ where $d$ is the distance between the atomic sites, where the DFT+DMFT band structure on a domain wall (which is a different type than ours) in commensurate CDW state has been reasonably well fitted\cite{PhysRevLett.122.036802}. We find that the tight-binding model reasonably fits well with the DFT+U data.

\begin{figure}[h]
\centering
\includegraphics[scale=0.5]{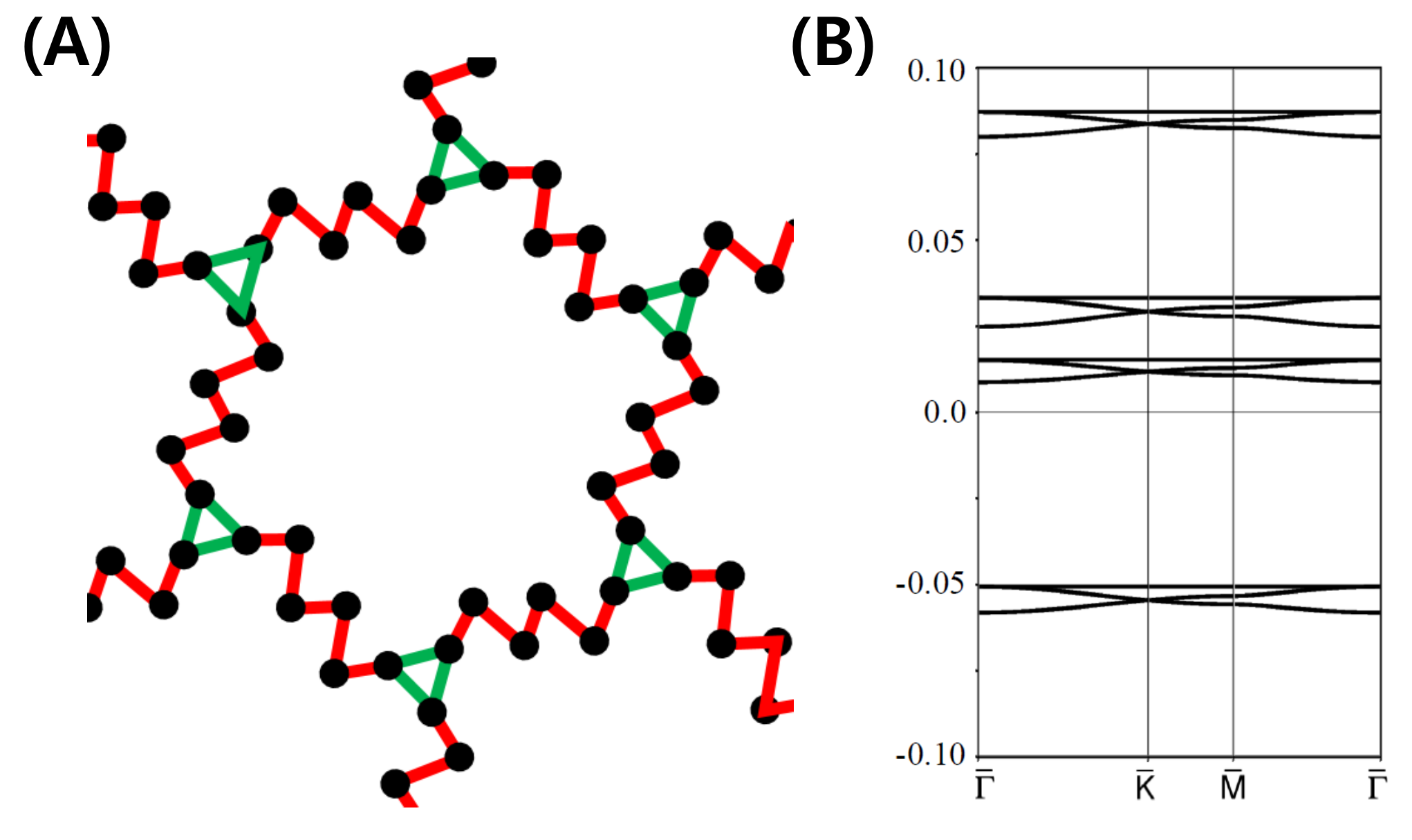}
\caption{(A) Geometry of network of NC-CDW 1T-TaS2. The black dots represent the Ta atomic sites inside the domain wall and junction of the network, which will contribute to the low-energy electronics. (B) Band structure near the Fermi level obtained from the tight-binding model in the network geometry. Y-axis is in the dimension of eV. The band structure features the cascade of flat bands as the models in main text. }
\label{DW-2}
\end{figure}

Following this, we find that the spectrum of the two models match reasonably well, see \ref{DW-1} (B). Moreover, using these tight-binding parameters, we can construct the tight-binding model for the whole network Fig.\ref{DW-2} (A), which clearly show the series of the flat bands Fig.\ref{DW-2} (B).



\bibliography{Network_0310}